\documentstyle[12pt]{article}
\def\np    { Nucl. Phys. }

\def\pl    { Phys. Lett. }

\def\beqa{\begin{eqnarray}}
\def\eeqa{\end{eqnarray}}
\def\parn              {  \par\noindent }
\def\parbigskip        {  \par\bigskip  }

\def\parbigskipn        {  \par\bigskip\noindent  }

\def\papertitlepage{\baselineskip 3.5ex \thispagestyle{empty}}
\def\Title#1{\vspace{1.5cm}\begin{center}
 {\Large\bf #1} \end{center} 
\vspace{1cm}}
\def\Authors#1{\begin{center} {\large\it #1} \end{center}}
\def\Abstract{\vspace{0.3cm}\begin{center} {\large\bf Abstract} 
           \end{center} \parbigskip}
\def\ICRRnumber#1#2#3{\hfill \begin{minipage}{3cm} #1
              \parn #2 \parn #3 \end{minipage}}
\begin{document}
\papertitlepage
\vspace*{-1 cm}
\ICRRnumber{ }{December 1998}{ }
\Title{Superalgebras in Many Types of \\
\vskip 1.5ex  M-Brane Backgrounds and\\
\vskip 1.5ex Various Supersymmetric Brane Configurations
} 
\Authors{{\sc\  Takeshi Sato
\footnote{tsato@icrr.u-tokyo.ac.jp}} \\
 \vskip 3ex
 Institute for Cosmic Ray Research, University of Tokyo, \\
3-2-1 Midori-cho, 
Tanashi, Tokyo 188-8502 Japan \\
}
\Abstract
We derive superalgebras in many types of supersymmetric 
M-brane backgrounds.
The backgrounds examined here include the cases of 
the M-wave
and the M-Kaluza-Klein monopole.
On the basis of the obtained algebras,
we deduce
all the supersymmetric non-orthogonal intersections of
the M-Kaluza-Klein monopole and the M-5-brane
at angles.
In addition,
we present a 1/4 supersymmetric 
worldvolume 3-brane soliton 
on the M-5-brane in 
the M-5-brane background as an extended solution of 
the 3-brane solitons of the M-5-brane by Howe, Lambert and West.
This soliton can be interpreted as a certain 
intersection of three M-5-branes.

\newpage

\section{Introduction}

The M-theory is currently a most hopeful candidate for a unified
theory of particle interactions\cite{tow2}\cite{wit1}
and is extensively studied from various points of
view\cite{tow3}\cite{banks1}\cite{sus1}. Among them,
the analysis via superalgebra is one of the most powerful approaches
to investigate its various properties\cite{hul1}\cite{tow7}. 
Since there are, of course, two kinds of supersymmetries,
two kinds of algebras have been discussed so far:
spacetime superalgebra and worldvolume ones.
The former was first constructed as the most general modification
of the standard D=11 supersymmetry algebra\cite{hol1}\cite{tow4},
and then 
deduced explicitly from M-brane actions in the {\it flat}
background via Noether method\cite{tow5}(see also ref.\cite{asc1}).
The latter, defined on the flat (p+1)-dimensional
worldvolumes of p-branes, 
were constructed as the 
maximal extensions of the (p+1)-dimensional
supertranslation algebras\cite{how1}\cite{tow6}.
Various possible supersymmetric brane 
intersections were deduced from both of the above algebras.
The same analyses were also applied to
the cases of string 
theories\cite{asc1}\cite{tow6}\cite{ham1}\cite{hatsu1},
although there are some
subtleties in the worldvolume cases.
In this way 
the discussions of superalgebras had been based
only on the {\it flat} cases until recently.

In the previous paper\cite{sato1}, however, we have proposed
the method of deriving spacetime superalgebras 
in supersymmetric brane backgrounds, i.e.{\it ``non-flat''} cases,
in terms of M-theory.\footnote{
The possibility of this computation has been pointed out
in the earlier paper ref.\cite{asc1} for a different
purpose (related to nontrivial topologies),
although it was not shown explicitly in it.}
The first motivation for this extension 
to non-flat cases has been 
to get the superalgebras of the 10-dimensional massive IIA theory
\cite{berg6}\cite{berg7}\cite{berg5},
which does not admit the flat background
owing to the existence of the cosmological 
constant\cite{rom1}\cite{berg3}.
(We have applied the method to this case in ref.\cite{sato2}.)
The idea presented in ref.\cite{sato1} is as follows:
let us consider a ``test'' brane,
the action of which is invariant under
local super-transformation.
First, suppose we take the background of the test brane 
to be 
flat, as done in ref.\cite{asc1}\cite{tow5}.
Then, the flat background has supertranslation symmetry, of course,
and the brane action is proved to have 
the same supertranslation symmetry.
In other words, 
{\it the symmetry of the brane action in a fixed background is
equal to
the (unbroken) symmetry of the fixed background}. 
So, we can define the corresponding Noether supercharge
and obtain the superalgebra.
Next, suppose the test brane
to be in a brane background 
which have some portions of supersymmetry.\footnote{
Here, we assume that this background actually consists of
such a large number of coincident M-branes
that the setting of ``test brane'' is justified,
as done in ref.\cite{tow8}.}
(This means that we take its background 
to be the brane solution.)
Then, by analogy with the above case,
the test brane action can be expected
to have the same portions of supersymmetry as the background. 
If the action does have the supersymmetry, 
it should be possible 
to define the corresponding Noether supercharge
and obtain the superalgebra in the same way.
(We call it ``the superalgebra via the brane probe''
because we ``probe'' the supersymmetry of the background
via the test brane.)
Since the anti-commutator of the supercharge 
is written in terms of an embedding of the test brane 
in the brane background, 
the consistency of this method should be confirmed
by deducing from the superalgebra
the previously obtained
supersymmetric configurations of two M-branes,
as the corresponding 
supersymmetric embeddings.\footnote{
In this case the test brane corresponds to 
one of the two M-branes and the background
corresponds to the other.}
In the paper\cite{sato1} we have examined the above
idea explicitly 
in the four cases:
a test M-2-brane and a test M-5-brane 
in the M-2-brane background and the M-5-brane background,
and we have confirmed
their consistency
by deducing from the algebras all the 1/4-supersymmetric
orthogonal intersections 
of the four combinations of
two M-branes known
before\cite{guv1}\cite{tow9}\cite{tsey2}\cite{gau1}\cite{tow6}. 

It is not evident, however, that the above discussions hold true 
in cases of backgrounds and probes other than the
M-2- and the M-5-branes.
So, the first purpose of this paper is
to clarify
{\it how generically the method is applicable}.
In section 2, we investigate all the cases of 1/2-supersymmetric
``basic'' M-brane backgrounds and probes possible to discuss:
the M-wave, the M-2-brane, the M-5-brane
and the M-Kaluza-Klein monopole
as backgrounds, and the M-wave, the M-2-brane and the M-5-brane
as probes (i.e. eight extra cases in addition to the previous 
four ones). 
The concrete procedures are as follows: First,
we substitute one of the M-brane solutions for the background of
each test brane action as was done
in ref.\cite{tow8}, and 
prove the invariance of the action under the unbroken supersymmetry
transformation. Next, we
derive the representation of the supercharge 
in terms of the worldvolume fields of the
test brane and their conjugate momenta,
compute its anti-commutator to obtain the superalgebra,
Finally, we confirm their consistency
by deducing {\it all} the previously known supersymmetric
orthogonal intersections of {\it any} combinations of the two
M-branes among the above.
We note that we cannot discuss the cases of the other
1/2-supersymmetric
``basic'' M-branes: the M-9-brane background,
the M-Kaluza-Klein monopole probe and the M-9-brane probe.
This is 
because full $\kappa$-symmetric actions have not been
constructed yet in these cases (see ref.\cite{loz1}\cite{berg9}).

Another purpose of this paper,
inspired by the above extension, is
to investigate
supersymmetric configurations in this set-up, i.e.
supersymmetric embeddings
of test branes in brane backgrounds. 
The supersymmetric configurations examined here include
not only 
{\it non-orthogonal intersections} of two M-branes at angles
but also {\it a nontrivial worldvolume soliton
on a brane in a brane background.}
As to the former cases, 
non-orthogonal intersections at angles have been 
investigated so far 
by using constraints of Killing spinors of the branes,
as in ref.\cite{berk1}\cite{kal1}\cite{ohta1}.
However,
it should be possible 
to deduce them
from the obtained superalgebras,
since
{\it all} the intersecting configurations of two branes
should be expressed as the 
corresponding embeddings of probes in brane
backgrounds in this method.
What we should do first is 
to find the supersymmetric embeddings that correspond to
the non-orthogonally intersecting two M-branes 
at angles,
and the next is to examine the preserved supersymmetry
for each value of the angles. 
In subsection 3.1 
we investigate the cases of the M-5-brane in 
the M-Kalza-Klein monopole background, and
deduce all the supersymmetric
intersections of the two M-branes at angles 
from the superalgebras,
most of which have been previously unexamined.

In subsection 3.2 we present the latter:
{\it a worldvolume 3-brane soliton on the M-5-brane in
the M-5-brane background}, that is,
this is a solution of the equations of motions of the brane action
in the {\it nontrivial} (brane) background,
while the usual worldvolume solitons are constructed 
as the solutions of the equations of motions of the brane action
in the {\it flat} 
backgrounds\cite{cal1}\cite{how3}\cite{gib1}\cite{how1}.
The soliton we present here 
is the extended version of the 3-brane soliton
of the M-5-brane presented by Howe, Lambert and West
in ref.\cite{how1}.
The soliton
is interesting not only in that
it can be interpreted as a certain intersection
of {\it three} M-5-branes,
but also in that each of the three branes is
expressed in a
way different from the other two:
a solution of supergravity, 
a(n) (embedded) source, 
and a worldvolume soliton.
We can easily prove by using the superalgebra
that the soliton has 1/4 supersymmetry.

Throughout this paper,
the invariance of the test brane actions are proved
to the {\it full} order in $\theta$,
while 
the explicit computations are performed only up to the low
orders which might contribute to the central charges 
at zeroth order in fermionic coordinates $\theta$.
(It is very difficult to derive
superalgebras to the {\it full} order in $\theta$.)
The obtained algebras, however, is useful
enough since we can discuss
all the supersymmetric configurations
only on the basis of the bosonic terms of the superalgebras.
The important fact in the computations is that
we can reduce the superspace in a brane background
with supercoordinates $(x,\theta)$ 
to that with the coordinates $(x,\theta^{+})$,
where the index $+$ of $\theta^{+}$ implies
that $\theta^{+}$ has a definite
worldvolume chirality of the background. 
The reason is the following: 
since half of supersymmetry is already not the symmetry of the system 
owing to the existence of the background brane,
the corresponding parameter $\theta^{-} $ must not be transformed.
So, the conjugate momentum of $\theta^{-} $ does not appear in the
supercharge $Q^{+}$, which means that the terms
including $\theta^{-} $ cannot contribute to the
central charges at zeroth order in $\theta$.
So, we ignore the terms including $\theta^{-}$ 
and {\it set $\theta^{-}=0$ from the beginning}.

This paper is organized as follows:
in section 2 we show the invariance of various test M-brane actions
in different M-brane backgrounds stated above
under the supertransformation corresponding to the symmetry
of the backgrounds,
and derive superalgebras in the backgrounds, respectively.
In subsection 2.1 we begin with the M-2-brane background, 
and we deal with the M-5-brane background
in subsection 2.2. 
We discuss the M-wave background in subsection 2.3, and
the M-Kalza-Klein monopole background in subsection 2.4.
In subsection 3.1 
we deduce all the non-orthogonal intersections
of the M-5-brane with
the M-Kaluza-Klein monopole at angles from the superalgebra. 
In subsection 3.2  we present a worldvolume 3-brane soliton
on the M-5-brane in the M-5-brane background.
Finally, in section 4 we give short summary. 

Before starting discussions we present the notations in this paper.
We use ``mostly plus'' metrics for both worldvolume and spacetime.
We use capital Latin letters($M,N,..$) for superspace indices, 
small Latin letters ($m,n,..$) for spacetime vectors 
and early small Greek letters
($\alpha,\beta$,..) for spinors.
Furthermore, we use late Greek letters     
($\mu,\nu,..$) for spacetime vectors
parallel to the background branes and
early Latin letters ($a,b,..$) for spacetime vectors transverse to them.
We use {\it hatted letters} ($\hat{M},\hat{m},\hat{a},\hat{\alpha}..$) 
for {\it all the inertial frame indices}
and middle Latin letters($i,j,..$) for worldvolume vectors.
And we use Majorana ($32 \times 32$) representation for Gamma matrices 
$\Gamma_{\hat{m}}$ which are all real and satisfy 
$ \{ \Gamma_{\hat{m}} , \Gamma_{\hat{n}} \} 
= 2\eta_{\hat{m}\hat{n}}$. 
$\Gamma_{\hat{0}}$ is antisymmetric and others 
symmetric. Charge Conjugation is ${\cal C}=\Gamma^{\hat{0}}$.
We denote the eleventh Gamma matrix $\Gamma_{\hat{10}}$ as
$\Gamma_{\hat{\natural}}$, as used in ref.\cite{tow7}.
Finally, we choose $\Gamma_{\hat{0}\hat{1}..\hat{9}\hat{\natural}}=1$. 


\section{Spacetime superalgebras in M-brane backgrounds}
\setcounter{footnote}{0} 
In this section we discuss spacetime superalgebras
in terms of various 1/2-supersymmetric M-brane backgrounds
and probes. 
In the following subsections
we discuss the M-2-brane, the M-5-brane, the M-wave and 
the M-Kaluza-Klein monopole background in turn, 
and probe each background
via the M-2-brane, the M-5-brane 
and the M-wave in this order.\footnote{
Though some of the combinations have been discussed
in ref.\cite{sato1}, we present them again
in order for this paper to be complete.} 
(This order is so arranged as to
begin with the easiest case to deal with and
discuss more difficult ones later.)
To be more concrete,
we first present the M-brane background solutions and
construct superfields and their supertransformations
in the backgrounds.
Then, we discuss superalgebras via each probe, separately. 
We prove the invariance of the test M-brane action under the
supertransformations which correspond to the symmetries of the
backgrounds,
derive spacetime superalgebras in the M-brane backgrounds
and deduce from them all the orthogonal intersections
the two M-branes
to confirm the consistency of this method.

\subsection{In the M-2-brane background}
First of all, we give some preliminaries about
the M-2-brane background, the superfields and
their supertransformation in the background.

The M-2-brane background solution is\cite{duf1}
\beqa
ds_{11}^{2} &=& H^{-2/3}\eta_{\mu\nu}dx^{\mu}dx^{\nu}
+H^{1/3}dy^{a}dy^{b}\delta_{ab}
\nonumber\\
C_{012}=H^{-1} &, &
\ \ \ \ {\rm (the \ others) }=0, \label{m2back}
\eeqa
where $\eta_{\mu\nu}$ is the 3-dimensional Minkovski metric with
coordinates $x^{\mu}$ and H is a harmonic function 
on the transverse 8-space with coordinates $y^{a}$, that is,
$H=1+\frac{q_{2}}{y^{6}}$ where $y=\sqrt{y^{a}y^{b}\delta_{ab}}$
and $q_{2}$ is a constant. 
This background admits a Killing spinor $\varepsilon$
which satisfies 
\beqa
\delta\psi_{m}
=(\partial_{m}+\frac{1}{4}\omega_{m}^{\ \hat{r}\hat{s}}
\Gamma_{\hat{r}\hat{s}}
+T_{m}^{\ n_{1}n_{2}n_{3}n_{4}}
F_{n_{1}n_{2}n_{3}n_{4}})\varepsilon =0
\eeqa
where $T_{m}^{\ n_{1}n_{2}n_{3}n_{4}}
=-\frac{1}{288}(\Gamma_{m}^{\ n_{1}n_{2}n_{3}n_{4}}
+8\Gamma^{ [ n_{1}n_{2}n_{3}}\delta_{m}^{n_{4} ] })$.
Then the Killing spinor has the form 
$ \varepsilon =H^{-1/6}\varepsilon_{0}$
where $\varepsilon_{0}$ has the  positive worldvolume chirality:
$\bar{\Gamma}\varepsilon_{0}\equiv \Gamma_{\hat{0}\hat{1}\hat{2}}
\varepsilon_{0}=+\varepsilon_{0}$.

Since $\bar{\Gamma}$ satisfies 
$\bar{\Gamma}^{T}=\bar{\Gamma}$ and $\bar{\Gamma}^{2}=1$,
both $\frac{1\pm\bar{\Gamma}}{2}$ and
$\frac{1\pm\bar{\Gamma}^{T}}{2}$ are projection operators.
So, if we denote 
$\ \frac{1\pm\bar{\Gamma}}{2}\zeta$ as $\zeta^{\pm}$
for a spinor $\zeta $,
the background 
is invariant under the transformation 
generated by the supercharge $Q^{+}$,
and each test brane action in this background
is also expected to be invariant.
On the other hand, the background and each brane action 
are not invariant under the transformation by $Q^{-}$,
which means that we should set the
corresponding transformation parameter
$\varepsilon^{-}$ to be zero.
Then, the conjugate momentum $\Pi^{-}$ 
of $\theta^{-}$ does not appear in the Noether charge $Q^{+}$. 
So, the terms including $\theta^{-}$ {\it never} contribute to 
the central charges at zeroth order in $\theta$.
Thus, we can ignore the terms and set $\theta^{-}=0$
from the beginning.
From now on, we will use these in all the cases we treat 
in this paper. Related with this,
we exhibit the properties of $\bar{\Gamma}$: 
\beqa
[ \bar{\Gamma},\Gamma_{\hat{\mu}} ] = [ \bar{\Gamma},{\cal C} ]
= \{ \bar{\Gamma},\Gamma_{\hat{a}}\} = 0. \label{gammabar}
\eeqa

Now, we have prepared to get the explicit representations of
the superfields and their supertransformations
in terms of superspace coordinates
to low orders in $\theta$.
By substituting (\ref{m2back}) and $\theta^{-}=0$
to the usual expressions\cite{cre1}
(and using (\ref{gammabar})),
we see that only the $E_{a}^{\ \hat{\alpha}}$ has 
the nontrivial contribution from the background.
From the results
the superspace 1-form on the inertial frame
$E^{\hat{A}}=dZ^{M}E_{M}^{\ \ \hat{A}}$ is given by
\footnote{In fact we need to know the (vanishing of the)
contribution from $E_{m}^{\ \hat{n}}$  
at order $\theta^{2}$. 
We can infer its vanishing 
in this specific simple background,    
but the expression of 
$E_{m}^{\ \hat{n}}$ at order $\theta^{2}$ in general background
was obtained
\cite{dew1}, 
by which our inference is confirmed.} 
\beqa
E^{\hat{\mu}} &=& dx^{\nu}H^{-1/3}\delta_{\nu}^{\ \hat{\mu}} 
-i\bar{\theta}^{+}\Gamma^{\hat{\mu}}d\theta^{+} +{\cal O}(\theta^{4})
\nonumber  \\
E^{\hat{a}} &=& dy^{b}H^{1/6}\delta_{b}^{\ \hat{a}}+{\cal
O}(\theta^{4})\nonumber  \\
E^{\hat{\alpha}} &=& d\theta^{\hat{\alpha}+}+\frac{1}{6}H^{-1}dH
\theta^{\hat{\alpha}+}+{\cal O}(\theta^{3})
\label{e1m2}.
\eeqa
Since the 1-form $E^{\hat{A}}$ has
no superspace (curved) indices,
$E^{\hat{A}}$
is invariant 
under the local supertransformation\cite{cre1} 
$\delta Z^{M}=\Xi^{M}$ in this background given by
\beqa
\Xi^{\mu} &=&i\bar{\varepsilon}^{+}\Gamma^{\mu}\theta^{+} 
+ {\cal O}(\theta^{3}) \nonumber \\
\Xi^{a} &=& 0+ {\cal O}(\theta^{3})  \nonumber \\
\Xi^{\alpha} &=& \varepsilon^{\alpha +}+ {\cal O}(\theta^{2}).
\label{supertr}
\eeqa
We can easily check the invariance of $E^{\hat{A}}$ explicitly
up to second order in $\theta$.
We note that 
the coordinates $y^{a}$ transverse to the background brane
are not transformed (at least up to the second order in $\theta $).
Namely, this is the supertranslation parallel to 
the background brane. (So, the Noether supercharge 
we will define later corresponds to this.)
It is also to be noted that
$\Gamma^{\mu}=H^{1/3}\Gamma^{\hat{\nu}}
\delta_{\hat{\nu}}^{\mu}$, i.e. the gamma matrices
with the spacetime
indices depend on the harmonic function.

The remaining fields are superspace gauge potentials:
3-form $C^{(3)}$ and 6-form $C^{(6)}$.
The former is introduced by the gauge invariant 4-form field
strength\cite{cre1}\cite{bri1}
\beqa
R^{(4)}\equiv dC^{(3)}=\frac{i}{2}E^{\hat{m}}E^{\hat{n}}
\bar{E^{\hat{\alpha}}}
(\Gamma_{\hat{n}\hat{m}})_{\hat{\alpha}\hat{\beta}}E^{\hat{\beta}}
+\frac{1}{4!}E^{\hat{m_{1}}}E^{\hat{m_{2}}}E^{\hat{m_{3}}}E^{\hat{m_{4}}}
F_{\hat{m_{4}}\hat{m_{3}}\hat{m_{2}}\hat{m_{1}}} \label{r4m2},
\eeqa
where $F_{\hat{m_{4}}\hat{m_{3}}\hat{m_{2}}\hat{m_{1}}}$ is the bosonic
field strength which is in this case 
associated with the electric M-2-brane background.

Here, we assume that
all the fermionic (but not bosonic) cocycles
in the superspaces 
are trivial.
(We assume this in all the cases in this paper.)
Then, the invariance of $R^{(4)}$ under
the super-transformation (\ref{supertr})
means that $\delta C^{(3)}$ 
can be written as a d-exact form to {\it full} 
order in $\theta$. 

From (\ref{r4m2}) we can get the explicit expression of $C^{(3)}$
as\footnote{Although the $\hat{\alpha}$ of $\theta^{\hat{\alpha}+}$
is the index of the inertial frame, 
$\theta^{\hat{\alpha}}
=\theta^{\beta}\delta_{\beta}^{\hat{\alpha}}+{\cal O}(\theta^{3}) $.
So, we need not distinguish the two indices in this paper.} 
\beqa
C^{(3)} &=& \frac{1}{3!}H^{-1}(-\epsilon_{\mu\nu\rho})
dx^{\mu}dx^{\nu}dx^{\rho}
-\frac{i}{2}H^{-2/3}dx^{\rho}\delta_{\rho}^{\hat{\mu}}
dx^{\sigma}\delta_{\sigma}^{\hat{\nu}}\bar
{\theta^{+}}\Gamma_{\hat{\mu}\hat{\nu}}d\theta^{+} \nonumber \\
& &
-\frac{i}{2}H^{1/3}dx^{c}\delta_{c}^{\hat{a}}
dx^{d}\delta_{d}^{\hat{b}}\bar
{\theta}^{+}\Gamma_{\hat{a}\hat{b}}d\theta^{+} 
\ \ + {\cal O}(\theta^{4}) \ \ \ \ (\epsilon_{012}=-1),
\label{c3m2}
\eeqa
and hence the supertransformation of $C^{(3)}$:
\beqa
\delta C^{(3)}\equiv  d(\bar{\varepsilon}^{+}\Delta_{2}) 
&=& d(-\frac{i}{2}H^{1/3}dy^{c}\delta_{c}^{\ \hat{a}}
dy^{d}\delta_{d}^{\ \hat{b}}\bar{\varepsilon}^{+}\Gamma_{\hat{a}\hat{b}}
\theta^{+}+ {\cal O}(\theta^{3})) \label{delc3}.
\eeqa

The latter superspace 6-form $C^{(6)}$ is introduced
by the 7-form field strength which takes the form\cite{leh1}  
\beqa
R^{(7)} & \equiv & dC^{(6)}-\frac{1}{2}C^{(3)}R^{(4)}\nonumber \\
& = & \frac{i}{5!}
E^{\hat{m_{1}}}...E^{\hat{m_{5}}}
\bar{E^{\hat{\alpha}}}
(\Gamma_{\hat{m_{5}}...\hat{m_{1}}})
^{\hat{\alpha}\hat{\beta}}E^{\hat{\beta}}
+\frac{1}{7!}E^{\hat{m_{1}}}...E^{\hat{m_{7}}}
F_{\hat{m_{7}}...\hat{m_{1}}}^{(7)}\label{r7m2}
\eeqa
where the 7-form $F^{(7)}$ is the Hodge dual of the bosonic 4-form
field strength.
We note that $C^{(6)}$ cannot be expressed globally in this case
because it has a part of magnetic potential 
which originates from the  existence of the M-2-brane.
(We denote it by $C^{(6)}_{\rm{mag}} $ formally.)
Then, in the same way as $\delta C^{(3)}$ 
the invariance of $R^{(7)}$ under (\ref{supertr}) 
means that $\delta C^{(6)}+\frac{1}{2}\delta C^{(3)}C^{(3)}$ 
can be written
as a d-exact form.\footnote{We note that $C^{(6)}_{\rm{mag}} $ is
invariant under the super-transformation (\ref{supertr})
owing to the inertness of the
transverse coordinates $y^{a}$ (see, (\ref{supertr})).}
From (\ref{r7m2}) we get
\beqa
\delta C^{(6)}+\frac{1}{2}\delta C^{(3)}C^{(3)}&\equiv& 
d(\bar{\varepsilon}^{+}\Delta_{5})\nonumber\\
=& &
d(-\frac{i}{4!}H^{1/3}dx^{\nu}\delta_{\nu}^{\mu} 
dy^{b_{1}}\delta_{b_{1}}^{\hat{a_{1}}}...
dy^{b_{4}}\delta_{b_{4}}^{\hat{a_{4}}}
(\Gamma_{\hat{a_{1}}...\hat{a_{4}}\mu}\theta^{+})^{\alpha}
+ {\cal O}(\theta^{3})).\ \ \ \label{sptrac6}
\eeqa
Now, we have finished preliminaries about the background,
the superfield and the supertransformation.
So, we will discuss each of the probes separately in the next.

\noindent
\underline{(2.1a)via the M-2-brane probe}

At first we review the case of a test M-2-brane floating
in the M-2-brane background discussed in ref.\cite{sato1}.
The M-2-brane action in a D=11 supergravity background is\cite{berg2} 
\beqa
S_{M2}=S^{(0)}+S_{WZ}=
-\int d^{3}\xi\sqrt{-{\rm det}\tilde{g}_{ij}}
+\int d^{3}\xi \frac{1}{3!}\varepsilon^{ijk}
\tilde{C}^{(3)}_{ijk}
\label{m2ea} 
\eeqa
where $\tilde{g}_{ij}
=E_{i}^{\hat{m}}E_{j}^{\hat{n}}\eta_{\hat{m}\hat{n}}$
is the induced worldvolume metric and $C^{(3)}_{ijk}$ is 
the worldvolume 3-form induced by the superspace 3-form gauge
potential.
$E_{i}^{\hat{A}}=\partial_{i}Z^{M}E_{M}^{\ \ \hat{A}}$
where $E_{M}^{\ \ \hat{A}}$ is the supervielbein. 
Note that the action is invariant under local 
super-transformation at this moment.
Let's fix the background to the M-2-brane solution (\ref{m2back}).
Since it holds $\delta E_{i}^{\hat{A}}=0$ under the 
supertransformation (\ref{supertr}), 
$ \tilde{g}_{ij}$ and hence $S^{(0)}$ are also invariant 
under (\ref{supertr}).  
On the other hand, from (\ref{delc3})
we get $\delta {\cal L}_{WZ}=d(\bar{\varepsilon}^{+}\Delta_{2})$.
So, the action (\ref{m2ea}) in the M-2-brane
background (\ref{m2back}) is invariant up to total derivative
under the supertransformation (\ref{supertr}).
(In the cases of the other backgrounds,
we can prove the invariance of the test M-2-brane action
under the supertransformations 
corresponding to the symmetries of
the backgrounds in the same way.)

So, we can define the corresponding Noether supercharge
$Q_{\alpha}^{+} $ in the Hamiltonian formulation 
as an integral over the test brane at fixed time
${\cal M}_{2}$, given by\cite{asc1}
\beqa
Q_{\alpha}^{+} &\equiv& Q_{\alpha}^{+(0)}-i\int_{{\cal M}_{2}} 
({\cal C}\Delta_{2})_{\alpha} \nonumber \\ 
&=&\int_{{\cal M}_{2}} d^{2}\xi(i\Pi_{\alpha}^{+}
-\Pi_{\mu}({\cal C}\Gamma^{\mu}\theta^{+})_{\alpha})
-\frac{1}{2}\int_{{\cal M}_{2}}
 dy^{a}dy^{b}({\cal C}\Gamma_{ab}\theta^{+})_{\alpha}
+{\cal O}(\theta^{3})
\eeqa
where $\Pi_{\mu}$ and $\Pi_{\alpha}^{+}$ are the conjugate momenta of 
$x^{\mu}$ and $\theta^{+}$, respectively. $Q_{\alpha}^{+(0)}$ is 
the momentum part,
the form of which is almost common to all the branes.
Then, we get the superalgebra of $Q_{\alpha}^{+}$ 
\beqa
\{ Q_{\alpha}^{+}, Q_{\beta}^{+}\}= 2\int_{{\cal M}_{2}}
d^{2}\xi\Pi_{\mu}
({\cal C}\Gamma^{\mu})_{\alpha\beta}
+\frac{2}{2}\int_{{\cal M}_{2}}  dy^{a}dy^{b}
({\cal C}\Gamma_{ab})_{\alpha\beta}
+{\cal O}(\theta^{2}).\label{algm2m2}
\eeqa
Before discussing this result,
we give the explicit expression of $\Pi_{\mu}$:
\beqa
\Pi_{\mu}&=&\frac{\delta {\cal L}^{(0)}}{\delta \dot{x}^{\mu}}
+\frac{1}{2} \varepsilon^{0ij}
\partial_{i}x^{\nu}\partial_{j}x^{\rho}C_{\mu\nu\rho}^{(3)}
+{\cal O}(\theta^{2}) 
\equiv \Pi_{\mu}^{(0)}+\Pi_{\mu}^{WZ}
\eeqa
where ${\cal L}^{(0)} $ is the Nambu-Goto Lagrangian.

The implications of this algebra are as follows:
since we are interested only in static configurations, 
we choose the static gauge:
\beqa
\partial_{0}x^{\mu}=\delta_{0}^{\ \mu},
\partial_{i}x^{0}=\delta_{i}^{\ 0}\label{static}.
\eeqa
Then, 
if the test brane is oriented parallel to the background brane,
the term like a central charge arises from the $\Pi_{\mu}^{WZ}$,
although the original central charge vanishes.
$\Pi_{\mu}^{(0)}$ and $\Pi_{\mu}^{WZ}$ are obtained respectively as
\beqa
\int_{{\cal M}_{2}} d^{2}\xi \Pi_{\mu}^{(0)} &=& 
|\int_{{\cal M}_{2}} dx^{1}dx^{2}H^{-1}|  
\delta_{\mu}^{\ 0}+{\cal O}(\theta^{2})\label{pi0} \\
\int_{{\cal M}_{2}} d^{2}\xi \Pi_{\mu}^{WZ}&=&
\int_{{\cal M}_{2}} dx^{1}dx^{2}H^{-1}
\delta_{\mu}^{\ 0}+{\cal O}(\theta^{2})\label{piwz}.
\eeqa
Thus, we conclude that the parallel configuration
(for example, $x^{1}=\xi^{1}, x^{2}=\xi^{2} $)
with 
a certain orientation
of the test brane has 1/2 spacetime supersymmetry
and the one with the other orientation breaks 
all the supersymmetry, which is consistent with
the previous result\cite{duf1}\cite{tsey1}.
We note that (\ref{pi0}) and (\ref{piwz})
are invariant under the 12-plane rotation and hence the discussion
above also holds, as it should be. 

On the other hand,
if the test brane is oriented orthogonally to the background brane
(i.e. zero-brane intersection),
the central charge {\it does} have the nonzero value.
In the static gauge with the test brane
to be fixed, for example, to 34-plane,
the algebra becomes
\beqa
\{ Q_{\alpha}^{+}, Q_{\beta}^{+}\}=2\int_{{\cal M}_{2}} 
d^{2}\xi H^{1/3}(1-\Gamma_{\hat{3}\hat{4}})_{\alpha\beta},
\eeqa
which means that 1/4 spacetime 
supersymmetry is preserved in this configuration (0$|$M2,M2).
We can easily see from the algebra (\ref{algm2m2})
that this is the only orthogonal 
intersection preserving supersymmetry,
which is also consistent with
the previous result given in
ref.\cite{guv1}\cite{tow9}\cite{tsey2}\cite{tow6}.

\noindent
\underline{(2.1b)via the M-5-brane probe}

The M-5-brane action is\cite{pst1}
\beqa
S_{M5}= S^{(0)} + S_{WZ}
&=& -\int d^{6}\xi [ 
\sqrt{-{\rm det}(g_{ij}+\tilde{{\cal H}}_{ij})}
+\frac{\sqrt{-g}}{4(\partial a)^{2}}(\partial_{i} a)
({\cal H})^{\ast ijk}{\cal H}_{jkl}(\partial^{l} a) ]
\nonumber\\
& &+\int (C^{(6)}+\frac{1}{2}{\cal H}C^{(3)}),\label{m5ea}
\eeqa
where ${\cal H}$ is the ``modified'' field strength
of the worldvolume self-dual 2-form ${\cal A}_{2}$
given by 
\beqa
{\cal H}=d{\cal A}_{2}-C^{(3)}.
\eeqa
$({\cal H})^{\ast ijk}$ and $\tilde{{\cal H}}^{ij}$ are
defined as
$({\cal H})^{\ast ijk}=\frac{1}{3!\sqrt{-g}}
\varepsilon^{ijki'j'k'}{\cal H}_{i'j'k'},\ 
\tilde{{\cal H}}^{ij}=\frac{1}{\sqrt{-(\partial a)^2}}
({\cal H})^{\ast ijk}\partial_{k}a$, respectively.
$a$ is an auxiliary worldvolume 
scalar field. 
The super-transformation of ${\cal A}_{2}$ is determined by
the requirement of the invariance of
the ``modified'' field strength ${\cal H}$ 
as in ref.\cite{tow5}. 
The transformation in this M-2-brane background is
\beqa
\delta {\cal A}_{2}=\bar{\varepsilon}^{+} \Delta_{2}
=-\frac{i}{2}H^{1/3}dy^{c}\delta_{c}^{\hat{a}}
 dy^{d}\delta_{d}^{\hat{b}}\bar{\varepsilon}^{+}
\Gamma_{\hat{a}\hat{b}}
\theta^{+}+{\cal O}(\theta^{3}),
\eeqa
where $\Delta_{2}$ is defined in (\ref{delc3}).
Since $g_{ij}$ and ${\cal H}$ are invariant,
the kinetic action $ S^{(0)}$ is also invariant under 
the transformation (\ref{supertr}).
On the other hand, $\delta {\cal L}_{WZ} $ is shown to be
the following d-exact form:
\beqa
\delta {\cal L}_{WZ}&=&\delta C^{(6)}
-\frac{1}{2}\delta C^{(3)}{\cal H}\nonumber \\
&=&d(\bar{\varepsilon}^{+}\Delta_{5}-\frac{1}{2}d{\cal A}_{2}
\bar{\varepsilon}^{+}\Delta_{2})\equiv 
d(\bar{\varepsilon}^{+}\Delta)\label{dl5inm2}
\eeqa
where $\bar{\varepsilon}^{+}\Delta_{5}$ is defined in (\ref{sptrac6}).
So, the M-5-brane action (\ref{m5ea}) in the M-2-brane
background is invariant up to total derivative
under the supertransformation (\ref{supertr}), and
the supercharge is given as before by
$Q_{\alpha}^{+} \equiv Q_{\alpha}^{+(0)}-i\int_{{\cal M}_{5}}
({\cal C}\Delta)_{\alpha} $
where $Q_{\alpha}^{+(0)} $ takes the form\cite{tow5} 
\beqa
Q_{\alpha}^{+(0)} =\int_{{\cal M}_{5}} d^{5}\xi [ i\Pi_{\alpha}^{+}
-\Pi_{\mu}({\cal C}\Gamma^{\mu}\theta^{+})_{\alpha}
+\frac{i}{2}{\cal P}^{\underline{i}\underline{j}}
({\cal C}(\Delta_{2})_{\underline{i}\underline{j}})_{\alpha} ],
\eeqa
where $\underline{i}$ is the space index of the test M-5-brane
worldvolume and $\Pi_{\mu}, \Pi_{\alpha}^{+}$ and 
${\cal P}^{\underline{i}\underline{j}}$ are the conjugate momenta of 
$x^{\mu}$, $\theta^{+}$ and 
${\cal A}_{\underline{i}\underline{j}}$, respectively.
Then, the superalgebra is obtained as
\beqa
\{ Q_{\alpha}^{+}, Q_{\beta}^{ +}\}= 2\int_{{\cal M}_{5}} d^{5}\xi
[ \Pi_{\mu}
({\cal C}\Gamma^{\mu})_{\alpha\beta} 
-\frac{1}{2}{\cal P}^{\underline{i}\underline{j}}
\partial_{\underline{i}}y^{a}
\partial_{\underline{j}}y^{a}
({\cal C}\Gamma_{ab})_{\alpha\beta} ] \nonumber \\
+ \frac{2}{4!}\int_{{\cal M}_{5}} 
dx^{\mu} dy^{a_{1}}...dy^{a_{4}}
({\cal C}\Gamma_{\mu a_{1}...a_{4}})_{\alpha\beta}
-\frac{2}{4}\int_{{\cal M}_{5}}
d{\cal A}_{2}dy^{a}dy^{b}
({\cal C}\Gamma_{ab})_{\alpha\beta}
+{\cal O}(\theta^{2}).\label{algm5m2}
\eeqa 
In the static gauge
the third term in (\ref{algm5m2}) means that
only the string intersection
with the M-2-brane background leads to preservation of
1/4 supersymmetry, which is again consistent
with ref.\cite{tow9}\cite{tsey2}\cite{tsey1}\cite{tow6}.

\noindent
\underline{(2.1c)via the M-wave probe}

The M-wave is a massless superparticle running at the the speed of
light. The action is\cite{bri2}
\beqa
S_{MW}=\int d\tau e(\tau)E_{\tau}^{\hat{m}}E_{\tau}^{\hat{n}}
\eta_{\hat{m}\hat{n}},\label{mwea}
\eeqa
where $\tau$ is the time on its worldline and $e(\tau)$ is an 
einbein.
Since $E_{\tau}^{\hat{M}}=\partial_{\tau} Z^{N} E_{N}^{\hat{M}}$
and $e(\tau)$ are invariant,
the action (\ref{mwea}) in the M-2-brane background
is also invariant under the supertransformation (\ref{supertr}).
We note that the action is exactly invariant because it has
no Wess-Zumino term.
Let us choose the gauge $e(\tau)=\frac{1}{2}$.
The supercharge is written as
\beqa
Q_{\alpha}^{+}=i\Pi_{\alpha}^{+}-\Pi_{\mu}({\cal C}
\Gamma^{\mu})_{\alpha},
\eeqa
and the superalgebra is obtained as
\beqa
\{ Q_{\alpha}^{+}, Q_{\beta}^{ +}\}= 2\Pi_{\mu}
({\cal C}\Gamma^{\mu})_{\alpha\beta}.\label{algmwm2}
\eeqa
We note that this form is exact to the full order in $\theta$.

Since $\Pi_{\mu}$ is the momentum parallel to the background
M-2-brane, supersymmetry is preserved only
if the absolute value of the parallel momentum
is equal to the energy $\Pi^{0}$.
For example,
when we fix the gauge (i.e. the embedding)
to
$x^{0}=x^{1}=\tau$, it holds $\Pi^{0}=\Pi^{1}$,
and the algebra is written as
\beqa
\{ Q_{\alpha}^{+}, Q_{\beta}^{ +}\}= 2\Pi^{0}
(1-\Gamma_{\hat{0}\hat{1}}),
\eeqa
which means that this embedding lead to preservation of
1/4 supersymmetry. Since this embedding can be interpreted as
(1$|$MW,M2) with 1/4 supersymmetry given in 
ref.\cite{tsey2}\cite{berg8}\cite{tow6}, 
the superalgebra (\ref{algmwm2})
is also consistent.
 
\subsection{In the M-5-brane background}

The M-5-brane background solution is given by \cite{guv1}
\beqa
ds_{11}^{2} &=& H^{-1/3}\eta_{\mu\nu}dx^{\mu}dx^{\nu}
+H^{2/3}dy^{a}dy^{b}\delta_{ab}
\nonumber\\
F_{abcd} &=& -\varepsilon_{abcde}\partial_{e} H ,\label{m5back}
\eeqa
where $\mu =0,1,..,5$ and $a=6,..,9,\natural $.
The Killing spinor $\varepsilon$ has the form
$\varepsilon =H^{-1/12}\varepsilon_{0}$
where $\varepsilon_{0}$ has the positive chirality of
the worldvolume of the background:
$\bar{\Gamma}'\varepsilon_{0}\equiv 
\Gamma_{\hat{0}\hat{1}\hat{2}\hat{3}\hat{4}\hat{5}}
\varepsilon_{0}=+\varepsilon_{0}$. 
Since $\frac{1\pm\bar{\Gamma}^{'(T)}}{2}$ are again 
projection operators,
we denote here
$\ \frac{1\pm\bar{\Gamma}'}{2}\zeta$ as $\zeta^{\pm}$
for a spinor $\zeta $.
Then, for the same reason 
stated in the case of the M-2-brane background,
only $Q^{+}$ is expected to be the symmetry of 
the system and   
we set $\varepsilon^{-}=0$ and hence $\theta^{-}=0 $.
We note that $\bar{\Gamma}'$ satisfies the (anti-)commutators
$ \{ \bar{\Gamma}', {\cal C} \} = 
\{ \bar{\Gamma}',\Gamma_{\hat{\mu}} \}=
 [ \bar{\Gamma}', \Gamma_{\hat{a}} ] = 0.$
By using this relations and the formula presented 
in ref.\cite{cre1},
the superspace 1-form on the inertial frame 
is given by
\beqa
E^{\hat{\mu}} &=& dx^{\nu}H^{-1/6}\delta_{\nu}^{\ \hat{\mu}} 
-i\bar{\theta}^{+}\Gamma^{\hat{\mu}}d\theta^{+} 
+{\cal O}(\theta^{4})\nonumber \\
E^{\hat{a}} &=& dy^{b}H^{1/3}\delta_{b}^{\ \hat{a}}+{\cal
O}(\theta^{4}) \nonumber \\
E^{\hat{\alpha}} &=& d\theta^{\hat{\alpha}+}+\frac{1}{12}H^{-1}dH
\theta^{\hat{\alpha}+}+{\cal O}(\theta^{3}).
\eeqa
The supertransformations of the supercoordinates
are the same forms as 
those in the M-2-brane background (\ref{supertr})
except for 
the ranges of $\mu$ and $a$. 
The superspace 3-form $C^{(3)}$ and the 6-form $C^{(6)}$
are introduced 
by (\ref{r4m2}) and (\ref{r7m2}), in the same way 
as the case of the M-2-brane background.
Note that 
$C^{(3)}$ cannot be expressed globally in this background 
because 
the 3-form has a magnetic part $C^{(3)}_{mag}$
which originate from the existence of the M-5-brane. 
However, $C^{(3)}_{mag}$ is invariant under the supertransformation
at least up to second order in $\theta$, 
owing to the inertness of the
transverse coordinates $y^{a}$ under the super-transformation.
As a result, 
\beqa
\delta C^{(3)}&\equiv& d(\bar{\varepsilon}^{+}\Delta_{2}')
=d(-iH^{1/6}dx^{\nu}\delta_{\nu}^{\hat{\mu}}
dy^{b}\delta_{b}^{\hat{a}}\bar{\varepsilon}^{+}
\Gamma_{\hat{\mu}\hat{a}}\theta^{+}+{\cal O}(\theta^{3}))
\label{trac3'}\\
\delta C^{(6)}+\frac{1}{2}\delta C^{(3)}C^{(3)}&\equiv& 
d(\bar{\varepsilon}^{+}\Delta_{5}')\nonumber \\
=& &
d(-\frac{i}{12}H^{1/3}dx^{\nu_{1}}\delta_{\nu_{1}}^{\mu_{1}}
...dx^{\nu_{3}}\delta_{\nu_{3}}^{\mu_{3}} 
dy^{b_{1}}\delta_{b_{1}}^{\hat{a_{1}}}
dy^{b_{2}}\delta_{b_{2}}^{\hat{a_{2}}}
(\Gamma_{\hat{\mu_{1}}...
\hat{\mu_{2}}
\hat{\mu_{3}}
\hat{a_{1}}\hat{a_{2}}}\theta^{+})^{\alpha}
\nonumber\\
& &-\frac{i}{4!}H^{1/3}dx^{\nu}\delta_{\nu}^{\mu} 
dy^{b_{1}}\delta_{b_{1}}^{\hat{a_{1}}}...
dy^{b_{4}}\delta_{b_{4}}^{\hat{a_{4}}}
(\Gamma_{\hat{a_{1}}...\hat{a_{4}}\mu}\theta^{+})^{\alpha}
\nonumber\\
& &-\frac{i}{2}C^{(3)}_{mag}dx^{\nu}\delta_{\nu}^{\mu}
dy^{b}\delta_{b}^{\hat{a}}
(\Gamma_{\hat{\mu}\hat{a}}\theta^{+})^{\alpha}
+ {\cal O}(\theta^{3}))\label{sptrac6m5}.
\eeqa
Since we have finished preliminaries about the M-5-brane background,
we will discuss each of the probes, respectively,
in the same way as the M-2-brane background.

\noindent
\underline{(2.2a)via the M-2-brane probe}

The test M-2-brane action is the same as (\ref{m2ea})
while the background is fixed 
to the M-5-brane solution (\ref{m5back}).
$ g_{ij}$ and hence $S^{(0)}$ are also invariant 
under the supertransformation for the same reason,
and the whole action (\ref{m2ea}) in the M-5-brane
background (\ref{m5back}) is invariant up to total derivative
because of (\ref{trac3'}).
So, we can define the corresponding Noether supercharge 
and obtain the superalgebra as
\beqa
\{ Q_{\alpha}^{+}, Q_{\beta}^{+}\}= 2\int_{{\cal M}_{2}}
d^{2}\xi\ \Pi_{\mu}
({\cal C}\Gamma^{\mu})_{\alpha\beta}
+2\int_{{\cal M}_{2}}dx^{\mu}dy^{a}
({\cal C}\Gamma_{\mu a})_{\alpha\beta}+{\cal O}(\theta^{2}).
\label{algm2m5}
\eeqa
The second term implies that
the string intersection of the test brane
with the background is the only 1/4-supersymmetric
configuration permitted in this background,
which is consistent with the previous results
\cite{tow9}\cite{tsey2}\cite{tsey1}\cite{tow6} and (3.1b).

\noindent
\underline{(2.2b)via the M-5-brane probe}

The test M-5-brane action is again (\ref{m5ea}) 
while the background is fixed as (\ref{m5back}).
The transformation of ${\cal A}_{2}$ in the M-5 brane background is 
determined by the invariance of ${\cal H}$,
just the same as the case of the M-2-brane background:
\beqa
\delta {\cal A}_{2}= \bar{\varepsilon}^{+}\Delta_{2}'=
-iH^{1/6}dx^{\nu}\delta_{\nu}^{\hat{\mu}}
dy^{b}\delta_{b}^{\hat{a}}\bar{\varepsilon}^{+}
\Gamma_{\hat{\mu}\hat{a}}
\theta^{+}+{\cal O}(\theta^{3})\label{da2inm5}.
\eeqa 
The proof of the invariance of the action in the M-5-brane
background is almost the same as that in the M-2-brane background,
except that $ \Delta_{5}$ and $\Delta_{2}$
in (\ref{dl5inm2}) 
are replaced with $ \Delta_{5}'$ and $\Delta_{2}'$ defined in 
(\ref{sptrac6m5}) and (\ref{trac3'}).
So, in the same way, the superalgebra in the M-5-brane background
is 
\beqa
\{ Q_{\alpha}^{+}, Q_{\beta}^{+}\}= 2\int_{{\cal M}_{5}} d^{5}\xi\ 
[ \Pi_{\mu}
({\cal C}\Gamma^{\mu})_{\alpha\beta} 
-\frac{1}{2}{\cal P}^{\underline{i}\underline{j}}
\partial_{\underline{i}}x^{\mu}
\partial_{\underline{j}}y^{a}
({\cal C}\Gamma_{\mu a})_{\alpha\beta} ]\nonumber \\
+\frac{2}{12}\int_{{\cal M}_{5}} 
dx^{\mu_{1}}...dx^{\mu_{3}} dy^{a_{1}}dy^{a_{2}}
({\cal C}\Gamma_{\mu_{1}...\mu_{3} a_{1}a_{2}})_{\alpha\beta}
+\frac{2}{4!}\int_{{\cal M}_{5}} 
dx^{\mu}dy^{a_{1}}...dy^{a_{4}}
({\cal C}\Gamma_{\mu a_{1}...a_{4}})_{\alpha\beta}\nonumber \\
-\frac{2}{2}\int_{{\cal M}_{5}}
(d{\cal A}_{2}-C^{(3)}_{mag})dx^{\mu}dy^{a}
({\cal C}\Gamma_{\mu a})_{\alpha\beta}
+{\cal O}(\theta^{2}),\label{algm5m5}
\eeqa
where $\Pi_{\mu}$ is
\beqa
\Pi_{\mu}&=&\frac{\delta {\cal L}^{(0)}}{\delta \dot{x}^{\mu}}
+\frac{1}{2} \varepsilon^{0i_{1}\cdots i_{5}}
\partial_{i_{1}}x^{\nu_{1}}\cdots
\partial_{i_{5}}x^{\nu_{5}}C_{\mu\nu_{1}\cdots\nu_{5}}^{(6)}
+{\cal O}(\theta^{2}),
\eeqa
where ${\cal L}^{(0)} $ is the kinetic term of the M-5-brane
Lagrangian.

The implications of the algebra are as follows:
in the static gauge (\ref{static})
the form of the momentum $\Pi_{\mu}$ is similar to that 
in the case of the test M-2-brane in the M-2-brane background.
So, a parallel configuration with a certain orientation 
leads to the preservation of 1/2 supersymmetry and the other
orientation breaks all the supersymmetry, 
which is consistent with the previous result\cite{guv1}\cite{tsey1}.
The third term implies that 1/4 supersymmetry is preserved
in the case of any three brane intersections.
Finally, 
we can prove that
any string intersections lead to
preservation of 1/4 supersymmetry.
This proof is a bit more complex than the others
because in addition to the fourth term,
the last term including the magnetic 3-form 
$ C^{(3)}_{mag}$ do not vanish in this case.
We will show that in the next.

Suppose ${\cal A}_{2}=0$, $\theta=0 $ and that
the test brane is fixed as 
$x^{1}=\xi^{1},y^{7}=\xi^{2},y^{8}=\xi^{3},
y^{9}=\xi^{4},y^{\natural}=\xi^{5}$.
Choosing the gauge $a=\xi^{0}$,
we have
\beqa
\{ Q_{\alpha}^{+}, Q_{\beta}^{ +}\}=2\int_{{\cal M}_{5}} d^{5}\xi\ 
[\Pi^{0}\cdot H^{-1/6}\delta_{\alpha\beta}
-H^{7/6}\{
{\cal C}\Gamma_{\hat{1}\hat{7}\hat{8}\hat{9}\hat{\natural}}
\nonumber\\
+H^{-1}( {\cal C}\Gamma_{\hat{1}\hat{7}}C_{mag89\natural}^{(3)}
+{\cal C}\Gamma_{\hat{1}\hat{8}}C_{mag79\natural}^{(3)}
+{\cal C}\Gamma_{\hat{1}\hat{9}}C_{mag78\natural}^{(3)}
+{\cal C}\Gamma_{\hat{1}\hat{\natural}}
C_{mag789}^{(3)}) \}_{\alpha\beta}]\label{algm5m5st}
\eeqa
where the momentum is given by
\beqa
\Pi^{0}=H^{4/3}\sqrt{1+H^{-2}[(C_{mag789}^{(3)})^{2}
+(C_{mag78\natural}^{(3)})^{2}
+(C_{mag79\natural}^{(3)})^{2}
+(C_{mag89\natural}^{(3)})^{2}]}\label{momm5m5st}
\eeqa
which originates from the kinetic term ${\cal L}^{(0)}$.
Then, since the last five matrices in (\ref{algm5m5st})
anti-commute with each other,
they can be gathered into a traceless matrix $\tilde{\Gamma}$
multiplied 
by their ``norm''
such as
\beqa
{\cal C}\Gamma_{\hat{1}\hat{7}\hat{8}\hat{9}\hat{\natural}}
+H^{-1}[{\cal C}\Gamma_{\hat{1}\hat{7}}C_{mag89\natural}^{(3)}
+{\cal C}\Gamma_{\hat{1}\hat{8}}C_{mag79\natural}^{(3)}
+{\cal C}\Gamma_{\hat{1}\hat{9}}C_{mag78\natural}^{(3)}
+{\cal C}\Gamma_{\hat{1}\hat{\natural}}
C_{mag789}^{(3)}]\nonumber\\
=H^{-4/3}\Pi^{0}\tilde{\Gamma},\label{normm5m5st}
\eeqa  
where $(\tilde{\Gamma})^{2}=1$.
So, 1/4 supersymmetry is preserved in this case, too.
We can see from the algebra (\ref{algm5m5})
that these are the only orthogonal intersections 
preserving supersymmetry.
All of the above are consistent with the result of 
ref.\cite{tow9}\cite{tsey2}\cite{gau1}\cite{berg8}\cite{tow6}.


\noindent
\underline{(2.2c)via the M-wave probe}

The action is (\ref{mwea}) while the 
background is fixed to the M-5-brane solution.
The proof of the invariance of the action in this background
is again the same as the one in the M-2-brane case.
So,  we only show the results.
The superalgebra is obtained as
\beqa
\{ Q_{\alpha}^{+}, Q_{\beta}^{ +}\}= 2\Pi_{\mu}
({\cal C}\Gamma^{\mu})_{\alpha\beta}.\label{algmwm5}
\eeqa
As is the case with the M-2-brane background,
1/4 supersymmetry is preserved
only in the embeddings in which
the absolute value of 
the momentum parallel to the background brane
is equal to the energy $\Pi^{0}$, which is also 
consistent with ref.\cite{tsey2}\cite{berg8}\cite{tow6}.

\subsection{In the M-wave background}

In this subsection we discuss the M-wave background,
which is given by\cite{hul2}
\beqa
ds^{2}=(K-1)dt^{2}-2Kdtdx^{1}+(1+K)(dx^{1})^{2}
+dy^{a}dy^{b}\delta_{ab},({\rm the\  others}=0)\label{mwback}
\eeqa
where K is a harmonic function in the variables 
$-t+x^{1}$ and $y^{a}$ ($a=2,3,..,\natural$).
The Killing spinor in this background is constant and
satisfies $\Gamma''\varepsilon \equiv 
\Gamma_{\hat{0}\hat{1}}\varepsilon=+\varepsilon$.
Since $\frac{1\pm\bar{\Gamma}^{''(T)}}{2}$ are again 
projection operators, we denote here
$\ \frac{1\pm\bar{\Gamma}''}{2}\zeta$ as $\zeta^{\pm}$
for a spinor $\zeta $.
Then, for the same reason as the case of the previous backgrounds, 
only $Q^{+}$ is the symmetry of 
the system and   
we set $\varepsilon^{-}=0$ and hence $\theta^{-}=0 $.
We note that $\bar{\Gamma}''$ satisfies the (anti-)commutators
$ \{ \bar{\Gamma}'', {\cal C} \} = 
\{ \bar{\Gamma}'',\Gamma_{\hat{\mu}} \}=
 [ \bar{\Gamma}'', \Gamma_{\hat{a}} ] = 0.$
By using these 
relations and the formula presented in ref.\cite{cre1},
the superspace 1-form on the inertial frame $E^{\hat{M}}
\equiv dZ^{N}E_{N}^{\hat{M}}$ is given by
\beqa
E^{\hat{0}}&=&(1-\frac{K}{2})dx^{0}+\frac{K}{2}dx^{1}-i
\bar{\theta}^{+}\Gamma^{\hat{0}}d\theta^{+}\nonumber\\
E^{\hat{1}}&=&-\frac{K}{2}dx^{0}+(1+\frac{K}{2})dx^{1}-i
\bar{\theta}^{+}\Gamma^{\hat{1}}d\theta^{+}\nonumber\\
E^{\hat{a}}&=&dy^{\hat{a}}\nonumber\\
E^{\hat{\alpha}}&=&d\theta^{+\hat{\alpha}}.
\eeqa
The supertransformations of the supercoordinates
are the same forms as 
those in the M-2-brane background (\ref{supertr}) except that
$\mu=0,1$ and $a=2,3,..9,\natural$. 
The superspace 3-form $C^{(3)}$ and the 6-form $C^{(6)}$
are introduced
by (\ref{r4m2}) and (\ref{r7m2}), in the same way 
as the case of the M-2-brane background.
Their (combinations of) supertransformations can also be written 
as d-exact forms by the same proofs, given by
\beqa
\delta C^{(3)}&\equiv& d(\bar{\varepsilon}^{+}\Delta_{2}'')
=d(-idx^{\mu}dy^{a}\bar{\varepsilon}^{+}
\Gamma_{\hat{\mu}\hat{a}}\theta^{+}+{\cal O}(\theta^{3}))
\label{trac3''}\\
\delta C^{(6)}+\frac{1}{2}\delta C^{(3)}C^{(3)}&\equiv& 
d(\bar{\varepsilon}^{+}\Delta_{5}'')=
d(-\frac{i}{4!}dx^{\mu}dx^{a_{1}}\cdots dx^{a_{4}}
\bar{\varepsilon}^{+}
\Gamma_{\hat{\mu}\hat{a_{1}}\cdots \hat{a_{4}}}
\theta^{+}+{\cal O}(\theta^{3})).\ \ \ \ \ 
\eeqa
We note that $\frac{1}{2}\delta C^{(3)}C^{(3)}$ is 
${\cal O}(\theta^{3})$ in this case.

Next we discuss each of the probes, respectively.
The original actions are the same as the previous cases and
the proofs of the invariance of the test brane actions
under the supertransformation
are also the same, while the background is replaced by
(\ref{mwback}).
So, we present only the results and their implications.

\noindent
\underline{(2.3a)via the M-2-brane probe}

The superalgebra is
\beqa
\{ Q_{\alpha}^{+}, Q_{\beta}^{+}\}= 2\int_{{\cal M}_{2}}
d^{2}\xi\ \Pi_{\mu}
({\cal C}\Gamma^{\mu})_{\alpha\beta}
+2\int_{{\cal M}_{2}}dx^{\mu}dy^{a}
({\cal C}\Gamma_{\mu a})_{\alpha\beta}+{\cal O}(\theta^{2}).
\label{algm2mw}
\eeqa
The second term implies that in the static gauge
only the string intersection leads to the
preservation of 1/4 supersymmetry, which is also 
consistent with ref.\cite{tsey2}\cite{berg8}\cite{tow6}.

\noindent
\underline{(2.3b)via the M-5-brane probe}

In this case the supertransformation of ${\cal A}_{2}$ is 
$\delta {\cal A}_{2}=\bar{\varepsilon}^{+}\Delta_{2}''$
where $\Delta_{2}''$ is defined in (\ref{trac3''}).
The superalgebra is
\beqa
\{ Q_{\alpha}^{+}, Q_{\beta}^{+}\}= 2\int_{{\cal M}_{5}} d^{5}\xi\ 
[ \Pi_{\mu}
({\cal C}\Gamma^{\mu})_{\alpha\beta} 
-\frac{1}{2}{\cal P}^{\underline{i}\underline{j}}
\partial_{\underline{i}}x^{\mu}
\partial_{\underline{j}}y^{a}
({\cal C}\Gamma_{\mu a})_{\alpha\beta} ]\nonumber \\
+\frac{2}{4!}\int_{{\cal M}_{5}} 
dx^{\mu}dy^{a_{1}}...dy^{a_{4}}
({\cal C}\Gamma_{\mu a_{1}...a_{4}})_{\alpha\beta}
-\frac{2}{2}\int_{{\cal M}_{5}}
d{\cal A}_{2}dx^{\mu}dy^{a}
({\cal C}\Gamma_{\mu a})_{\alpha\beta}
+{\cal O}(\theta^{2})\label{algm5mw}
\eeqa
The third term implies that only string intersection leads to the
preservation of 1/4 supersymmetry, which is also 
consistent with ref.\cite{tsey2}\cite{berg8}\cite{tow6}.

\noindent
\underline{(2.3c)via the M-wave probe}

In the gauge $e(\tau)=\frac{1}{2}$
the momentum is given by
\beqa
\Pi_{0}&=&-(1-K)\partial_{\tau}x^{0}-K \partial_{\tau}x^{1}
+{\cal O}(\theta^{2})\nonumber\\
\Pi_{1}&=&-K\partial_{\tau}x^{0}+(1+K)\partial_{\tau}x^{1}
+{\cal O}(\theta^{2})\nonumber\\
\Pi_{a}&=&\partial_{\tau}y^{a}.
\eeqa
Then, the superalgebra becomes 
\beqa
\{ Q_{\alpha}^{+}, Q_{\beta}^{ +}\}= 2\Pi_{\mu}
({\cal C}\Gamma^{\mu})_{\alpha\beta}=2(\partial_{\tau}x^{0}
-\partial_{\tau}x^{1}),
\label{algmwmw}
\eeqa
which means that only such an embedding as $x^{0}=x^{1}=k\tau$ 
for a positive constant $k$
leads to 
preservation of 1/2 supersymmetry.
This is the configuration (1$|$MW,MW) with 1/2 supersymmetry,
which is also consistent with ref.\cite{berg8}\cite{tow6}.

\subsection{In the M-Kaluza-Klein monopole background}

The M-Kaluza-Klein monopole background solution is \cite{tow2}
\beqa
ds_{11}^{2}=\eta_{\mu\nu}dx^{\mu}dx^{\nu}+Vdy_{a}dy_{b}\delta^{ab}
+V^{-1}(dy^{\natural}-A_{a}dy^{a})^{2}\label{mkback}
\eeqa
where $\mu,\nu = 0,1..6 $ and $ a,b = 7,8,9$.
$\eta_{\mu\nu}$ is the 7-dimensional Minkovski metric with
coordinates $x^{\mu}$. 
$A_{a}$ is a magnetic potential of a monopole
on the transverse 3-space with coordinates $y^{a}$
and $V$ is a harmonic function on the same 3-space
satisfying the equation: 
$\partial_{a}V=\varepsilon^{abc}\partial_{b}A_{c}$.

This background admits a constant Killing spinor $\varepsilon$
which satisfies $\bar{\Gamma}'''\varepsilon_{0}\equiv 
\Gamma_{\hat{0}\hat{1}..\hat{6}}
\varepsilon_{0}=+\varepsilon_{0}$. 
For the same reason, if we denote
$\ \frac{1\pm\bar{\Gamma}'''}{2}\zeta$ as $\zeta^{\pm}$
for a spinor $\zeta $, $Q^{+}$ corresponds to the symmetry
of the system, while $Q^{-}$ is not,
and we set $\varepsilon^{-}=0$ and $\theta^{-}=0 $
from the beginning. $\bar{\Gamma}'''$ satisfies the relations
$[ \bar{\Gamma}''',{\cal C} ]=[ \bar{\Gamma}''',\Gamma_{\hat{\mu}} ]
= \{ \bar{\Gamma}''',\Gamma_{\hat{a}}\} 
= \{ \bar{\Gamma}''',\Gamma_{\hat{\natural}}\}=0$.

The superspace 1-form on the inertial frame 
is given by
\beqa
E^{\hat{\mu}} &=& dx^{\nu}\delta_{\nu}^{\ \hat{\mu}} 
-i\bar{\theta}^{+}\Gamma^{\hat{\mu}}d\theta^{+} 
+{\cal O}(\theta^{4})\nonumber \\
E^{\hat{a}} &=& V^{1/2}dy^{b}\delta_{b}^{\ \hat{a}}
+{\cal O}(\theta^{4}) \nonumber \\
E^{\hat{\natural}} &=& V^{-1/2}dy^{\natural}- V^{-1/2}A_{a}dy^{a}
+{\cal O}(\theta^{4}) \nonumber \\
E^{\hat{\alpha}} &=&d\theta^{\hat{\alpha}+}+{\cal O}(\theta^{3}).
\eeqa
The supertransformations of the supercoordinates
are again the same forms as 
those in the M-2-brane background (\ref{supertr})
except that
$\mu=0,1,..,6$ and $a=7,..,9,\natural$.
The (combinations of) supertransformations
of the superspace 3-form $C^{(3)}$ and the 6-form $C^{(6)}$
in this background are proved to be
d-exact forms given respectively by
\beqa
\delta C^{(3)}&\equiv&d(\bar{\varepsilon}^{+}\Delta_{2}''')
=d(-\frac{i}{2}dx^{\mu_{1}}\delta_{\mu_{1}}^{\hat{\nu_{1}}} 
dx^{\mu_{2}}\delta_{\mu_{2}}^{\hat{\nu_{2}}} 
\bar{\varepsilon}^{+}
\Gamma_{\hat{\nu_{1}}\hat{\nu_{2}}}\theta^{+}\nonumber\\
&-&\frac{i}{2}Vdy^{a_{1}}\delta_{a_{1}}^{\hat{b_{1}}}
dy^{a_{2}}\delta_{a_{2}}^{\hat{b_{2}}}
\bar{\varepsilon}^{+}
\Gamma_{\hat{b_{1}}\hat{b_{2}}}\theta^{+}\nonumber\\
& &-idy^{a}\delta_{a}^{\hat{b}}dy^{\natural}\bar{\varepsilon}^{+}
\Gamma_{\hat{b}\hat{\natural}}\theta^{+}
-iA_{a}dy^{a}dy^{b}\delta_{b}^{\hat{c}}\bar{\varepsilon}^{+}
\Gamma_{\hat{c}\hat{\natural}}\theta^{+}
+{\cal O}(\theta^{3}))\label{trac3'''}\\
\delta C^{(6)}+\frac{1}{2}\delta C^{(3)}C^{(3)}&\equiv& 
d(\bar{\varepsilon}^{+}\Delta_{5}''')=
d(-\frac{i}{5!}dx^{1}\cdots dx^{5}\bar{\varepsilon}^{+}
\Gamma_{\hat{\mu_{1}}\cdots \hat{\mu_{5}}}\theta^{+}\nonumber\\
&-&\frac{i}{12}Vdx^{1}\cdots dx^{3}dy^{a}dy^{b}\bar{\varepsilon}^{+}
\Gamma_{\hat{\mu_{1}}\cdots \hat{\mu_{3}}\hat{a}\hat{b}}
\theta^{+}\nonumber\\
&-&\frac{i}{3!}dx^{1}\cdots dx^{3}dy^{a}dy^{\natural}
\bar{\varepsilon}^{+}
\Gamma_{\hat{\mu_{1}}\cdots \hat{\mu_{3}}\hat{a}\hat{\natural}}
\theta^{+}\nonumber\\
&-&\frac{i}{3!}Vdx^{\mu}dy^{a_{1}}\cdots dy^{a_{3}}dy^{\natural}
\Gamma_{\hat{\mu} \hat{a_{1}}\cdots \hat{a_{3}}\hat{\natural}}
\theta^{+}\nonumber\\
&-&\frac{i}{3!}dx^{1}\cdots dx^{3}A_{a}dy^{a}dy^{b}
\bar{\varepsilon}^{+}
\Gamma_{\hat{\mu_{1}}\cdots \hat{\mu_{3}}\hat{b}\hat{\natural}}
\theta^{+}
+{\cal O}(\theta^{3})).\label{trac6'''}
\eeqa

Next, we will discuss each of the probes, respectively.
Each original action is the same as the previous cases 
while the background is chosen to be (\ref{mkback}).
Since the proofs of the invariance of the test brane actions
under the supertransformation
are also the same, 
we present only the results and their implications again.

\noindent
\underline{(2.4a)via the M-2-brane probe}

The superalgebra in the M-Kaluza-Klein monopole background
via the M-2-brane probe is
\beqa
\{ Q_{\alpha}^{+}, Q_{\beta}^{+}\}= 2\int_{{\cal M}_{2}}
d^{2}\xi\ \Pi_{\mu}
({\cal C}\Gamma^{\mu})_{\alpha\beta}
+\frac{2}{2}\int_{{\cal M}_{2}}dx^{\mu}dx^{\nu}
({\cal C}\Gamma_{\hat{\mu} \hat{\nu}})_{\alpha\beta}
+\frac{2}{2}\int_{{\cal M}_{2}}Vdy^{a}dy^{b}
({\cal C}\Gamma_{\hat{a} \hat{b}})_{\alpha\beta}\nonumber \\
+2\int_{{\cal M}_{2}}dy^{a}dy^{\natural}
({\cal C}\Gamma_{\hat{a} \hat{\natural}})_{\alpha\beta}
+2\int_{{\cal M}_{2}}A_{a}dy^{a}dy^{b}
({\cal C}\Gamma_{\hat{b} \hat{\natural}})_{\alpha\beta}
+{\cal O}(\theta^{2})\label{spalgm2mk}.\ \ \ \ \ 
\eeqa
In the static gauge
the second term implies that 1/4 supersymmetry is
preserved in the case of 2-brane intersection.
The last three terms imply that 1/4 supersymmetry is also
preserved in 0-brane intersection.
The proof of this preservation is essentially similar to
the proof in the case of string intersection 
of the two M-5-branes in (2.2b).
If the test brane is fixed as $y^{a}=\xi^{1},y^{\natural}=\xi^{2}$,
only the fourth term in addition to the first term does not vanish.
So, we can easily see that 1/4 supersymmetry is preserved.
But, if the test brane is fixed as, for example,
$y^{7}=\xi^{1},y^{8}=\xi^{2}$,
it is not so simple because
the last term do contribute to the r.h.s. of the algebra,
which becomes
\beqa
\{ Q_{\alpha}^{+}, Q_{\beta}^{ +}\}=2\int_{{\cal M}_{2}}d^{2}\xi\ [
\sqrt{V^{2}+(A_{7})^{2}+(A_{8})^{2}}\cdot 1
+V{\cal C}\Gamma_{\hat{7}\hat{8}}+A_{7}{\cal C}
\Gamma_{\hat{8}\hat{\natural}}
-A_{8}{\cal C}\Gamma_{\hat{7}\hat{\natural}}
]_{\alpha\beta}.\label{m2mk0int}
\eeqa
Then, since the last three matrices in (\ref{m2mk0int})
anti-commute with each other,
they can be gathered into a traceless matrix $\tilde\Gamma'$
($(\tilde\Gamma')^{2}=1$)
multiplied 
by their ``norm'', which is equal to the energy 
$\Pi^{0}=\sqrt{V^{2}+(A_{7})^{2}+(A_{8})^{2}}$.
So, 1/4 supersymmetry is
preserved in this embedding.
We can see again from the algebra (\ref{spalgm2mk})
that these are the only orthogonal intersections 
preserving supersymmetry.
All of the above are consistent with the result of 
ref.\cite{tsey2}\cite{berg8}\cite{tow6}.


\noindent
\underline{(2.4b)via the M-5-brane probe}

The superalgebra in the M-Kaluza-Klein monopole background
via the M-5-brane probe is
\beqa
\{ Q_{\alpha}^{+}, Q_{\beta}^{+}\}= 2\int_{{\cal M}_{5}} d^{5}\xi
\ [ \Pi_{\mu}
({\cal C}\Gamma^{\mu})_{\alpha\beta} 
-\frac{1}{2}{\cal P}^{\underline{i}\underline{j}}
\partial_{\underline{i}}x^{\mu}
\partial_{\underline{j}}x^{\nu}
({\cal C}\Gamma_{\hat{\mu}\hat{\nu}})_{\alpha\beta} 
] \nonumber \\
+\frac{2}{5!}\int_{{\cal M}_{5}}dx^{\mu_{1}}...dx^{\mu_{5}}
({\cal C}\Gamma_{\hat{\mu_{1}}...\hat{\mu_{5}}})_{\alpha\beta}
+\frac{2}{12}\int_{{\cal M}_{5}} 
Vdx^{\mu_{1}}...dx^{\mu_{3}} dy^{a_{1}}dy^{a_{2}}
({\cal C}\Gamma_{\hat{\mu_{1}}...\hat{\mu_{3}} 
\hat{a_{1}}\hat{a_{2}}})_{\alpha\beta}
\nonumber \\
+\frac{2}{3!}\int_{{\cal M}_{5}} 
dx^{\mu_{1}}...dx^{\mu_{3}} dy^{a}dy^{\natural}
({\cal C}\Gamma_{\hat{\mu_{1}}...\hat{\mu_{3}}
\hat{a} \hat{\natural}})_{\alpha\beta}
+\frac{2}{3!}\int_{{\cal M}_{5}} 
dx^{\mu_{1}}...dx^{\mu_{3}} A_{a}dy^{a}dy^{b}
({\cal C}\Gamma_{\hat{\mu_{1}}...\hat{\mu_{3}} 
\hat{b} \hat{\natural}})_{\alpha\beta}
\nonumber \\
+\frac{2}{3!}\int_{{\cal M}_{5}} 
Vdx^{\mu}dy^{a_{1}}...dy^{a_{3}}dy^{\natural}
({\cal C}\Gamma_{\hat{\mu} 
\hat{a_{1}}...\hat{a_{4}}})_{\alpha\beta}
-\frac{2}{2}\int_{{\cal M}_{5}}
d{\cal A}_{2}
[dx^{\mu}dx^{\nu}({\cal C}\Gamma_{\hat{\mu}\hat{\nu}})_{\alpha\beta}
\nonumber \\
+Vdy^{a}dy^{b}({\cal C}\Gamma_{\hat{a}\hat{b}})_{\alpha\beta}
+dy^{a}dy^{\natural}({\cal C}
\Gamma_{\hat{a}\hat{\natural}})_{\alpha\beta}
+A_{a}dy^{a}dy^{b}({\cal C}
\Gamma_{\hat{b}\hat{\natural}})_{\alpha\beta}
+{\cal O}(\theta^{2})].\ \ \ \ \label{algm5mk}
\eeqa
The third term implies that 5-brane intersection
leads to preservation of
1/4 supersymmetry. The fourth, the fifth and the sixth terms
implies that 3-brane intersection
also leads to preservation of
1/4 supersymmetry.
We can prove the latter by the same procedure as the case of
the zero-brane intersection of the test M-2-brane with this
M-KK background in (2.4c).
The seventh term means that 1/4 supersymmetry is
also preserved in the string intersection.
We can see from the algebra that there are no other orthogonal
intersections with supersymmetry.
All of the above are again consistent with the result of 
ref.\cite{tsey2}\cite{berg8}\cite{tow6}.

\noindent
\underline{(2.4c)via the M-wave probe}

The superalgebra in the M-Kaluza-Klein monopole background
via the M-wave probe is
\beqa
\{ Q_{\alpha}^{+}, Q_{\beta}^{ +}\}= 2\Pi_{\mu}
({\cal C}\Gamma^{\mu})_{\alpha\beta}.\label{algmwmk}
\eeqa
Since there are no central charges and 
the momentum $\Pi^{\mu}$ is parallel to the background,
(1$|$MW,MKK) is the only embeddings to preserve (1/4)
supersymmetry, which is also 
consistent with ref.\cite{tsey2}\cite{berg8}\cite{tow6}. 

\section{Various supersymmetric brane configurations from 
the superalgebras}
\setcounter{footnote}{0} 

\subsection{Supersymmetric intersections of the M-5-brane with
the M-Kaluza-Klein monopole at angles
from the superalgebra}

In this subsection we discuss non-orthogonal
supersymmetric intersections
of two M-branes at angles.
In the same way 
as the paper ref.\cite{berk1}\cite{kal1}\cite{ohta1}, 
we start from the configurations of test branes 
``maximally parallel'' to background branes and rotate the  
test branes.
When either  
the background or the probe is the M-2-branes or the M-waves,
at most only two angles are needed to parametrize
generic rotation from ``maximally parallel'' configurations.
Then, the combinations have supersymmetry 
only in the cases of orthogonal
intersections, all of which are already known
and has been reproduced in the previous section.
So, the only combinations of two M-branes discussed above
to permit the existence of  
non-orthogonal intersections with supersymmetry 
are the cases of (3.2b) and (3.4b):
the test M-5-brane in the M-5-brane background
and
the test M-5-brane in the M-Kaluza-Klein monopole background.
(The former is investigated in detail in ref.\cite{ohta1}.)
When the two branes
intersect orthogonally, we can prove straightforwardly
the preservation of 1/4 supersymmetry in both cases.
When they intersect non-orthogonally at angles, however,
the right hand side of the superalgebra (\ref{algm5m5})
in the former case (3.2b) becomes
very complicate, especially owing to the existence of 
the magnetic 3-form gauge potential $C^{(3)}_{mag}$, and
it is difficult to see how much supersymmetry is preserved.
So, in this paper 
we discuss non-orthogonal intersections in the latter case (3.4b). 
Although the superalgebra (\ref{algm5mk}) also
appears to include the magnetic 1-form $A_{a}$,
the algebra can be written
such that $A_{a}$ does not appear in it,
in the expression of only vector (i.e. completely ``hatless'') 
indices, as
\beqa
\{ Q_{\alpha}^{+}, Q_{\beta}^{+}\}= 2\int_{{\cal M}_{5}} d^{5}\xi
\Pi_{\mu}
({\cal C}\Gamma^{\mu})_{\alpha\beta} 
+{\cal P}^{\underline{i}\underline{j}}
\{ -\frac{1}{2}
\partial_{\underline{i}}x^{\mu}
\partial_{\underline{j}}x^{\nu}
({\cal C}\Gamma_{\mu\nu})_{\alpha\beta} 
\} \nonumber \\
+\frac{2}{5!}\int_{{\cal M}_{5}}dx^{\mu_{1}}...dx^{\mu_{5}}
({\cal C}\Gamma_{\mu_{1}...\mu_{5}})_{\alpha\beta}
+\frac{2}{12}\int_{{\cal M}_{5}} 
dx^{\mu_{1}}...dx^{\mu_{3}} dy^{a_{1}}dy^{a_{2}}
({\cal C}\Gamma_{\mu_{1}...\mu_{3} 
a_{1}a_{2}})_{\alpha\beta}
\nonumber \\
+\frac{2}{3!}\int_{{\cal M}_{5}} 
dx^{\mu_{1}}...dx^{\mu_{3}} dy^{a}dy^{\natural}
({\cal C}\Gamma_{\mu_{1}...\mu_{3}
a \natural})_{\alpha\beta}
+\frac{2}{3!}\int_{{\cal M}_{5}} 
dx^{\mu}dy^{a_{1}}...dy^{a_{3}}dy^{\natural}
({\cal C}\Gamma_{\mu a_{1}...a_{4}})_{\alpha\beta}\nonumber\\
-\frac{2}{2}\int_{{\cal M}_{5}}
d{\cal A}_{2}
[dx^{\mu}dx^{\nu}({\cal C}\Gamma_{\mu\nu})_{\alpha\beta}
+dy^{a}dy^{b}({\cal C}\Gamma_{ab})_{\alpha\beta}
+dy^{a}dy^{\natural}({\cal C}
\Gamma_{a\natural})_{\alpha\beta}]
+{\cal O}(\theta^{2}).\label{algm5mk'}
\eeqa
We investigate the preserved supersymmetry on the basis of this 
expression.

Let us find the embeddings corresponding to
the intersection of the M-5-brane with
the M-Kaluza-Klein monopole at angles.
Since the M-Kaluza-Klein monopole is essentially a 6-brane,
the generic rotation is parametrized by four independent angles 
$\theta_{i}$ ($i=1,..,4$). (Namely,
they always intersect at least on a string.)
A ``naive'' embedding is expected without loss of generality
as
$x^{0}=\xi^{0},
x^{i}=\xi^{i}\cos\theta_{i}, y^{i+6}=\xi^{i}\sin\theta_{i}
(i=1,2,3,4),\  x^{5}=\xi^{5}$ and $x^{6}={\rm const}$,
where $\theta_{i}$ is the ``angle'' of the rotation 
of the i-(i+6)-plane for $i=1,2,3,4$, respectively.
This embedding, however, does not preserve any supersymmetry
in the cases of non-orthogonal intersections.
Instead, the supersymmetry is preserved by the embedding:
\beqa
\frac{\partial x^{\mu}}{\partial \xi^{i}}=\cos\theta_{i}
e_{\hat{i}}^{\mu}=\cos\theta_{i}
\delta_{i}^{\mu}& &{\rm \ \ \ (for \ \ i=1,..,4) },\nonumber\\
\frac{\partial y^{a}}{\partial \xi^{i}}=\sin\theta_{i}
e_{\hat{i+6}}^{a}& &{\rm \ \ \ (for \ \ i=1,..,3)},\nonumber\\
\frac{\partial y^{a}}{\partial \xi^{4}}=\sin\theta_{4}
e_{\hat{\natural}}^{a}
&,&\frac{\partial x^{\mu}}{\partial
\xi^{5}}=e_{\hat{5}}^{\mu}=\delta_{i}^{\mu},
\frac{\partial x^{6}}{\partial
\xi^{i}}=0\label{4angleembed'}
\eeqa
where $e_{\hat{m}}^{n}$ is the inverse of a vielbein 
$e_{m}^{\ \hat{n}}$
in the M-Kaluza-Klein monopole background (\ref{mkback}).
This is so constructed as to satisfy the relations:
\beqa
\frac{ds^{2}(y(\xi^{i}+d\xi^{i});y(\xi^{i}))}
{ds^{2}(x(\xi^{i}+d\xi^{i});x(\xi^{i}))}
=\tan^{2}\theta_{i} {\rm \ \ \ (for\ \  i=1,2,3,4)}.
\eeqa
Namely, (\ref{4angleembed'}) is the embedding to 
{\it keep
each intersecting angle $\theta_{i}$ fixed in terms of 
the square of
infinitesimal distance $ds^{2}$}. 
In fact this embedding makes 
the induced worldvolume metric to be the (1+5)-dimensional 
flat one, i.e. $g_{ij}\equiv
\frac{\partial x^{m}}{\partial \xi^{i}}
\frac{\partial x^{n}}{\partial \xi^{j}}g_{mn}=\eta_{ij}$.
In other words, the test brane is embedded so as to keep its flatness.
\footnote{
Although this embedding is not written explicitly
but written in terms of differential equations,
the embedding can be determined by 
integrating them from points at infinity.}

When the test M-5-brane is embedded as (\ref{4angleembed'}),
the superalgebra (\ref{algm5mk'}) becomes\footnote{
We omit hats of
the vector indices of Gamma matrices from now on.}
\beqa 
\{ Q_{\alpha}^{+}, Q_{\beta}^{+} \}&=&2\int_{{\cal M}_{5}}d\xi^{5}(1
-\cos\theta_{1}\cos\theta_{2}\cos\theta_{3}\cos\theta_{4}
\Gamma_{012345}\nonumber\\
&-&\sin\theta_{1}\sin\theta_{2}\cos\theta_{3}\cos\theta_{4}
\Gamma_{078345}
-\sin\theta_{1}\cos\theta_{2}\sin\theta_{3}\cos\theta_{4}
\Gamma_{072845}\nonumber\\
&-&\sin\theta_{1}\cos\theta_{2}\cos\theta_{3}\sin\theta_{4}
\Gamma_{0723\natural 5}
-\cos\theta_{1}\sin\theta_{2}\sin\theta_{3}\cos\theta_{4}
\Gamma_{018945}\nonumber\\
&-&\cos\theta_{1}\sin\theta_{2}\cos\theta_{3}\sin\theta_{4}
\Gamma_{0183\natural 5}
-\cos\theta_{1}\cos\theta_{2}\sin\theta_{3}\sin\theta_{4}
\Gamma_{0129\natural 5}\nonumber\\
&-&\sin\theta_{1}\sin\theta_{2}\sin\theta_{3}\sin\theta_{4}
\Gamma_{0789\natural 5}).\label{algm5mkang'}
\eeqa
Since the gamma matrix products in (\ref{algm5mkang'})
commute with each other and $\Gamma_{0123456}$,
all these matrices can be simultaneously diagonalized.
So, we can analyse the above consequence 
by the same technique as that in ref.\cite{ohta1}.
Their eigenvalues are all $\pm 1$
because the square of
them are all equal to the identity.
And since all of them
are traceless,
we can arrange for the following five
matrices to be such a basis as 
\beqa
\Gamma_{0123456}=\bar{\Gamma}_{A}&\equiv&
{\rm diag.}(\overbrace{1,\cdots,1} ^{\mbox{16}} 
,\overbrace{-1,\cdots,-1} ^{\mbox{16}}),\nonumber\\
\Gamma_{012345}=\bar{\Gamma}_{B}&\equiv&
{\rm diag.}(\overbrace{1,\cdots,1} ^{\mbox{8}} 
,\overbrace{-1,\cdots,-1} ^{\mbox{8}},\cdots),\nonumber\\
\Gamma_{078345}=\bar{\Gamma}_{C}&\equiv&
{\rm diag.}(\overbrace{1,\cdots,1} ^{\mbox{4}} 
,\overbrace{-1,\cdots,-1} ^{\mbox{4}},
\overbrace{1,\cdots,1} ^{\mbox{4}} 
,\overbrace{-1,\cdots,-1} ^{\mbox{4}},\cdots), \nonumber\\
\Gamma_{072845}=\bar{\Gamma}_{D}&\equiv&
{\rm diag.}(1,1,-1,-1,1,1,-1,-1,1,1,-1,-1,1,1,-1,-1,\cdots),
\nonumber\\
\Gamma_{0723\natural 5}=\bar{\Gamma}_{E}&\equiv&
{\rm diag.}(1,-1,1,-1,1,-1,1,-1,1,-1,1,-1,1,-1,1,-1,\cdots).
\label{gammadiag}
\eeqa
The representations of
the rest of the matrices appearing in (\ref{algm5mkang'})
are determined because each of the rest is the product of 
the above five. 
(We note that it is sufficient for us to know only
the first 16 components of the 
matrices because $ Q^{+}$ is the supercharge projected by the 
matrix $\frac{1+\bar{\Gamma}_{A}}{2}$.)
Then, we can derive the following expression:
\beqa
\{ Q_{\alpha}^{+}, Q_{\beta}^{+} \}=4\int_{{\cal M}_{5}}d\xi^{5}
\cdot {\rm diag.}
( \sin^{2}\frac{\theta_{1}-\theta_{2}-\theta_{3}-\theta_{4}}{2},
\sin^{2}\frac{\theta_{1}-\theta_{2}-\theta_{3}+\theta_{4}}{2},
\nonumber\\
\sin^{2}\frac{\theta_{1}-\theta_{2}+\theta_{3}-\theta_{4}}{2},
\sin^{2}\frac{\theta_{1}-\theta_{2}+\theta_{3}+\theta_{4}}{2},
\sin^{2}\frac{\theta_{1}+\theta_{2}-\theta_{3}-\theta_{4}}{2},
\nonumber\\
\sin^{2}\frac{\theta_{1}+\theta_{2}-\theta_{3}+\theta_{4}}{2},
\sin^{2}\frac{\theta_{1}+\theta_{2}+\theta_{3}-\theta_{4}}{2},
\sin^{2}\frac{\theta_{1}+\theta_{2}+\theta_{3}+\theta_{4}}{2},
\nonumber\\
\cos^{2}\frac{\theta_{1}+\theta_{2}+\theta_{3}+\theta_{4}}{2},
\cos^{2}\frac{\theta_{1}+\theta_{2}+\theta_{3}-\theta_{4}}{2},
\cos^{2}\frac{\theta_{1}+\theta_{2}-\theta_{3}+\theta_{4}}{2},
\nonumber\\
\cos^{2}\frac{\theta_{1}+\theta_{2}-\theta_{3}-\theta_{4}}{2},
\cos^{2}\frac{\theta_{1}-\theta_{2}+\theta_{3}+\theta_{4}}{2},
\cos^{2}\frac{\theta_{1}-\theta_{2}+\theta_{3}-\theta_{4}}{2},
\nonumber\\
\cos^{2}\frac{\theta_{1}-\theta_{2}-\theta_{3}+\theta_{4}}{2},
\cos^{2}\frac{\theta_{1}-\theta_{2}-\theta_{3}-\theta_{4}}{2}..).
\label{algm5mkang2}
\eeqa
We use this result to provide a systematic analysis of preserved
supersymmetry.
Before analyzing the result,
we clarify the ranges of $\theta_{i}$. 
As opposed to the cases of two M-branes of the same kind,
there are no differences between parallel ($\theta=0$) and 
``anti-parallel'' ($\theta=\pi$) configurations 
as to the combinations of two M-branes of different kinds.
So, we can set $-\frac{\pi}{2}\leq\theta_{i}\leq \frac{\pi}{2}$
without the loss of generality.


\noindent
\underline{(3a) one angle}

To begin with, we deal with the simplest case of a rotation
by single angle $\theta_{1}$, that is, 
we set the other angles to zero.
Then, denoting ${\bf 1_{n}}$ as the $n\times n$ identity matrices,
we get
\beqa 
\{ Q_{\alpha}^{+}, Q_{\beta}^{+} \}= 4\int_{{\cal M}_{5}}d\xi^{5}
 {\rm diag.}(
\sin^{2}
\frac{\theta_{1}}{2}{\bf 1_{8}},
\cos^{2}
\frac{\theta_{1}}{2}{\bf 1_{8}},\cdots),
\eeqa 
which means that all the supersymmetry is broken unless
$\theta_{1}=0$ (or $\theta_{1}=\pi$). When this condition is  
satisfied (i.e. (5$|$M5,MKK) given in ref.\cite{tsey2}),
1/4 supersymmetry is preserved.


\noindent
\underline{(3b) two angles}

Now, we get
\beqa 
\{ Q_{\alpha}^{+}, Q_{\beta}^{+} \}= 4\int_{{\cal M}_{5}}d\xi^{5}
 {\rm diag.}(
\sin^{2}
\frac{\theta_{1}-\theta_{2}}{2}{\bf 1_{4}},
\sin^{2}
\frac{\theta_{1}+\theta_{2}}{2}{\bf 1_{4}},\nonumber\\
\cos^{2}
\frac{\theta_{1}+\theta_{2}}{2}{\bf 1_{4}},
\cos^{2}
\frac{\theta_{1}-\theta_{2}}{2}{\bf 1_{4}},\cdots),
\eeqa 
which means that all the supersymmetry is broken unless
$\theta_{1}\pm \theta_{2}=0, \pm \pi$. 
When one of these conditions is  
satisfied, 1/8 supersymmetry is preserved.
When two of them are satisfied, 
1/4 supersymmetry is preserved, which are the cases of
$\theta_{1}=\pm\theta_{2}=\pm \frac{\pi}{2}$
(i.e. (3$|$M5,MKK)).

\noindent
\underline{(3c) three angles}

We have
\beqa 
\{ Q_{\alpha}^{+}, Q_{\beta}^{+} \}&=& 4\int_{{\cal M}_{5}}d\xi^{5}
 {\rm diag.}(
\sin^{2}
\frac{\theta_{1}-\theta_{2}-\theta_{3}}{2}{\bf 1_{2}},
\sin^{2}
\frac{\theta_{1}-\theta_{2}+\theta_{3}}{2}{\bf 1_{2}},
\sin^{2}
\frac{\theta_{1}+\theta_{2}-\theta_{3}}{2}{\bf 1_{2}},\nonumber\\
& &\sin^{2}
\frac{\theta_{1}+\theta_{2}+\theta_{3}}{2}{\bf 1_{2}},
\cos^{2}
\frac{\theta_{1}+\theta_{2}+\theta_{3}}{2}{\bf 1_{2}},
\cos^{2}
\frac{\theta_{1}+\theta_{2}-\theta_{3}}{2}{\bf 1_{2}},\nonumber\\
& &\cos^{2}
\frac{\theta_{1}-\theta_{2}+\theta_{3}}{2}{\bf 1_{2}},
\cos^{2}
\frac{\theta_{1}-\theta_{2}-\theta_{3}}{2}{\bf 1_{2}},\cdots),
\eeqa 
which means that all the supersymmetry is broken unless
$\theta_{1}\pm \theta_{2}\pm \theta_{3}=0, \pm \pi$.
When one of these conditions is  
satisfied, 1/16 supersymmetry is preserved.
In special cases the supersymmetry is enhanced.
When one of $\theta_{i}$ is $\pm\frac{\pi}{2}$ in addition to
the condition,
1/8 supersymmetry is preserved.
If another $\theta_{i}$ is $\pm\frac{\pi}{2}$,
1/4 supersymmetry is preserved, to be sure, but in this case
it holds the other angle is equal to zero or $\pm\pi$.
So, these should be classified in ``two angles''.

\noindent
\underline{(3d) four angles}

The superalgebra is given in (\ref{algm5mkang2}).
Supersymmetry is completely broken unless 
$\theta_{1}\pm \theta_{2}\pm \theta_{3}\pm \theta_{4}=0, \pm \pi$.
When one of these conditions is  
satisfied, 1/32 supersymmetry is preserved.
Furthermore, 
the supersymmetry is enhanced in the following cases:
suppose none of the indices i,j,k,l are the same.
Then, 1/16 supersymmetry is preserved
in the three cases:
\beqa
\left\{
  \begin{array}{@{\,}lll} 
\theta^{i}=\pm\theta^{j}&{\rm \ and \ \ \ \ }&
\theta^{k}=\pm\theta^{l}\\
\theta^{i}\pm\theta^{j}=\pm \frac{\pi}{2}& {\rm \ and \ \ \ \ }& 
\theta^{k}\pm\theta^{l}=\pm \frac{\pi}{2}\\
\theta^{i}\pm \frac{\pi}{2} & {\rm \ and \ \ \ \ }&
\theta_{i}\pm \theta_{j}\pm \theta_{k}=\pm\frac{\pi}{2}.
 \end{array}
\right.
\eeqa
3/32 supersymmetry is preserved in the two cases:
\beqa
\left\{
  \begin{array}{@{\,}lll} 
&\theta^{i}=\pm\theta^{j}=\pm\theta^{k}=\pm\theta^{l}& \\
\theta^{i}=\pm\theta^{j},&
\theta^{i}\pm\theta^{k}=\pm \frac{\pi}{2}{\rm \ \ \ and }&
\theta^{i}\pm\theta^{l}=\pm \frac{\pi}{2}.
\end{array}
\right.
\eeqa
1/8 supersymmetry is preserved
in the two cases:
\beqa
\left\{
  \begin{array}{@{\,}ll} 
\theta^{i}=\pm\theta^{j}=\pm\theta^{k}=\pm\theta^{l}
&=\pm \frac{\pi}{4} \\
\theta^{i}=\pm\theta^{j} {\rm \ \ \ and }&
\theta^{k}=\pm\theta^{k}=\pm \frac{\pi}{2}.
\end{array}
\right.
\eeqa
Finally, 1/4 supersymmetry is preserved
in the cases of $\theta^{i}=\pm\theta^{j}
=\pm\theta^{k}=\pm\theta^{l}=\pm \frac{\pi}{2}$,
which is (1$|$M5,MKK) given in ref.\cite{berg8}\cite{tow6}. 

\subsection{A worldvolume soliton on the M-5-brane in the
M-5-brane background}

We have discussed so far 
supersymmetric configurations of test branes
in (nontrivial) brane backgrounds so far.
In fact these configurations can be regarded as trivial
examples of worldvolume solitons on test branes
in brane backgrounds because they are solutions of
the equations of motion
of the test branes in the brane backgrounds.
In this subsection, related to this interpretation, 
we present a (nontrivial) supersymmetric
worldvolume 3-brane soliton 
on the (test) M-5-brane in a
{\it non-flat} (M-5-brane) background. 
Namely, this is a solution of
the equations of motion
of the M-5-branes in the M-5-brane background,
while usual worldvolume solitons
are constructed as the solutions of
branewave equations
of branes in {\it flat} 
backgrounds\cite{cal1}\cite{how3}\cite{gib1}\cite{how1}.
The soliton we present here is an extended solution of
3-brane solitons on the M-5-brane in the flat spacetime 
given by Howe, Lambert and West in ref.\cite{how1}.
So, before presenting our original result,
we give a short review of
the 3-brane solitons given in ref.\cite{how1}.

Suppose the M-5-brane lies in D=11 flat background
with coordinates $X^{m}$ (for m=0,1,..,9,$\natural$).
We denote the 6-dimensional worldvolume coordinates of it
by $\xi^{i}$ (i=0,1,..5), and suppose 
the 3-brane soliton lies in the hyperplane 
$\xi^{0},\xi^{1},\xi^{2},\xi^{3}$,
which means that
the coordinates transverse to the 3-brane
is $\xi^{4}$ and $\xi^{5}$, which we denote as $\xi^{i'}$.
Here, we assume that all fields depend only on the 
transverse coordinates $\xi^{i'}$ 
just like the cases of brane solutions of spacetimes.
Then,
the 3-brane soliton in ref.\cite{how1} (in our notation) is
written as
\beqa
X^{0}=\xi^{0},X^{1}=\xi^{1},X^{2}=\xi^{2},
X^{3}=\xi^{3},X^{4}=\xi^{4},X^{5}=\xi^{5},\nonumber\\
X^{6}=X^{6}(\xi^{4},\xi^{5}), 
X^{7}=q\ln{\sqrt{(\xi^{4})^{2}+(\xi^{5})^{2}}}\nonumber\\
X^{8},X^{9},X^{\natural} ={\rm (const.)},{\cal A}_{2}=0,
\label{wvsolitonflat}
\eeqa
where $X^{6}$ and $X^{7}$ satisfy the relations:
\beqa
\epsilon_{i'j'}\partial^{j'}X^{6}=\pm\partial_{i'}X^{7}
\eeqa
(where $\epsilon_{i'j'}$ is the volume element on the transverse space).
The scalar $X^{6}$ is interpreted as the
worldvolume dual of a 5-form field strength $G_{5}$ 
of a 4-form gauge field ${\cal B}_{4}$
defined as
\beqa
G_{i_{1}\cdots i_{5}}\equiv d{\cal B}_{4}=
\epsilon_{i_{1}\cdots i_{6}}\partial^{i_{6}}X^{6}.
\eeqa
So, $q$ is an electric charge in terms of the 4-form ${\cal B}_{4}$
while it is a magnetic charge in terms of $ X^{1}$.
So, $ X^{1}$ cannot be expressed globally,
though {\it the solution does exist}.
This is a worldvolume 3-brane soliton on the M-5-brane
with 1/2 worldvolume (i.e. 1/4 spacetime)
supersymmetry.
In terms of spacetime 
this soliton is interpreted as
the configuration of two M-5-branes intersecting on the 3-brane.
The 3-brane is considered to be 
an intersecting subspace of {\it another} M-5-brane, and
the two nontrivial scalars $X^{1},X^{2}$ pointing  
to the two directions transverse to the original M-5-brane, 
are considered to be the rest subspace of the another M-5-brane.

Now, we present our original result.
On the analogy of the above 3-brane soliton,
We construct by rule of thumb
a 3-brane soliton on the M-5-brane in the M-5-brane
background (\ref{m5back}) given by
\beqa
x^{0}=\xi^{0},x^{1}=\xi^{1},x^{2}=\xi^{2},x^{3}=\xi^{3},
y^{6}=\xi^{4},y^{7}=\xi^{5},\nonumber\\
x^{4}=x^{4}(\xi^{4},\xi^{5}),
x^{5}=q\ln\sqrt{(\xi^{4})^{2}+(\xi^{5})^{2}},\nonumber\\
y^{8},y^{9},y^{\natural}
={\rm constant},
{\cal A}_{2}=0,\label{wvsoliton1}
\eeqa
where $x^{4}$ is interpreted as 
the dual scalar of a worldvolume five-form field strength
$G_{5}$ defined as $G_{5}=(\partial x^{4})^{\star}$, and
$q$ is a magnetic charge of $x^{4}$.
$x^{4}$ and $x^{5}$ satisfy the equation: 
\beqa
\epsilon_{ij}\partial^{j}x^{4}=\partial_{i}x^{5},\ \ \ \ \ \
{\rm (for \ \ \ \ i',j'=4,5)}\label{45relation}
\eeqa
where $\epsilon_{i'j'}$ is the volume element on the
space $\xi^{4},\xi^{5} $. 

Let us first explain the interpretation of this soliton.
From upper part of (\ref{wvsoliton1})
the test brane is considered to be
embedded in the 12367-hyperplane, which means that
this is the
3-brane intersection of the test M-5-brane with the background
M-5-brane.
Besides, from the worldvolume point of view 
this can be interpreted as a 3-brane soliton in the plane 
$\xi^{0},\xi^{1},\xi^{2},\xi^{3}$,
while the coordinates transverse to the 3-brane
are $\xi^{4},\xi^{5}$.
And the two worldvolume scalars  point to $x^{4},x^{5}$.
So,
on the basis of the interpretation presented above,
this soliton can be interpreted as
the following three M-5-brane intersection: 
\begin{center}
\begin{tabular}{llllllcrrrrr}
directions &0 &1&2&3&4&5&6&7&8&9&$\natural$\\
the background M-5-brane& $\circ$ &$\circ$ & $\circ$ & 
$\circ$ & $\circ$ & $\circ$ & - & - & - & - & - \\
the test M-5-brane & $\circ$ & $\circ$ & $\circ$ & $\circ$ 
& - & - & $\circ$ & $\circ$ & - & - & - \\
the M-5-brane as the soliton 
& $\circ$ & $\circ$ & $\circ$ & $\circ$ & 
$\underline{\circ}$ & $\underline{\circ}$ & - & - & - & - & -
\end{tabular}
\end{center}
where the underbarred circles in the table mean 
that they are scalars transverse to the test M-5-brane.

Next, we show that the embedding (\ref{wvsoliton1}) is
a worldvolume soliton, i.e. a solution of 
the equations of motion
of the M-5-branes in the M-5-brane
background(\ref{m5back}).
To see this
we present the expressions of induced worldvolume fields
in this case:
the metric $\tilde{g}_{ij}$, the 3-form $\tilde{C}^{(3)}$,
and the 6-form $\tilde{C}^{(6)}$.
By using the relations (\ref{45relation}),
the induced worldvolume ($6\times 6$)
metric $\tilde{g}_{ij}$ is
\beqa
\tilde{g}_{ij}&=&
\left(
\begin{array}{ccc}
-H^{-1/3}&0&0\\
0&H^{-1/3}\cdot{\bf 1_{3\times3}}&0\\
0&0&H^{2/3}\cdot \delta_{i'j'}+H^{-1/3}
(\partial_{i'} x^{4}\partial_{j'} x^{4}
+\partial_{i'} x^{5}\partial_{j'} x^{5})
\end{array}
\right)\nonumber\\
&=&H^{-1/3}{\rm diag.}(-1,1,1,1,
H+\delta^{i'j'}\partial_{i'}x^{5}\partial_{j'}x^{5},
H+\delta^{i'j'}\partial_{i'}x^{5}\partial_{j'}x^{5}).
\eeqa
We note that ${\cal H}|_{\theta=0}=-\tilde{C}^{(3)}|_{\theta=0}=
-\tilde{C}_{mag}^{3}=0$
in the cases of
three brane intersection of the probe with the background,
because
$\tilde{C}_{mag}^{(3)}\neq 0$ only if the three indices are all
transverse to the background M-5-brane.
On the other hand, $\tilde{C}^{(6)}|_{\theta=0}=
\tilde{C}_{ele}^{(6)}$ has a only nonzero component
\beqa
\tilde{C}_{ele\ \ 012345}^{(6)}=
H^{-1}(\partial_{4}x^{4}
\partial_{5}x^{5}-\partial_{4}x^{5}
\partial_{5}x^{4})=
H^{-1}\delta^{i'j'}\partial_{i'}x^{5}\partial_{j'}x^{5}.
\eeqa
Then, we can see that
this soliton (\ref{wvsoliton1}) solves the equations of motion
of the M-5-branes (\ref{m5ea}) in the {\it nontrivial} (M-5-brane)
background (\ref{m5back}) given in this case by
\beqa
\frac{\delta{\cal L}}{\delta x^{m}}&=&
\partial_{i}[\frac{\delta{\cal L}}{\delta \partial_{i}x^{m}}]
\nonumber\\
&=&\partial_{i} 
[-\sqrt{-{\rm det } \tilde{g}}\tilde{g}^{ij}\partial_{i}
x^{n}g_{nm}+\frac{1}{5!}
\epsilon^{i\ i_{1}\cdots i_{5}}
\epsilon_{m\  m_{1}\cdots m_{5}}
\partial_{i_{1}}x^{m_{1}}\cdots
\partial_{i_{5}}x^{m_{5}}H^{-1}],\ \ \ \ \ 
\eeqa
where the last term is the contribution of $C_{ele}^{(6)}$
in the Wess-Zumino-like term of the M-5-brane action (\ref{m5ea}).
Thus, we can call it a worldvolume soliton.

The final issue we have to discuss is
the preserved supersymmetry of the soliton.
This configuration is expected to have
1/4 supersymmetry because it can be interpreted as the above 
three M-5-brane intersection.
We can easily confirm that 1/4 supersymmetry is preserved
by using the superalgebra (\ref{algm5m5}).
Substituting the solution (\ref{wvsoliton1}) for (\ref{algm5m5})
we have
\beqa
\{ Q_{\alpha}^{+}, Q_{\beta}^{+}\}= 2\int_{{\cal M}_{5}} d^{5}\xi
\delta_{\alpha\beta} +2\int_{{\cal M}_{5}} d^{5}\xi
({\cal C}\Gamma_{12367})_{\alpha\beta}.
\eeqa
So, 1/4 supersymmetry is preserved
in this solution.

Though it seems possible to construct 
other solutions of this type,
we do not discuss them here.

\section{Summary}

We have derived superalgebras in many type of M-brane backgrounds
via various probes, and checked their consistency by deducing
from the algebras all the previously known
supersymmetric orthogonal intersections
(and parallel configurations) of various combinations
of two M-branes.
In addition, on the basis of the superalgebras,
we have derived all the non-orthogonal
supersymmetric intersections of the M-5-brane
and the M-Kaluza-Klein monopole at angles,
most of which were previously unexamined.
Finally, we have presented
the 1/4-supersymmetric worldvolume 3-brane soliton
on the M-5-brane in the M-5-brane background,
which can be interpreted as the intersection
of three M-5-branes.

We note that the setup of this method is inappropriate
just on the background branes because of the singular
behavior of their metrics. Generally speaking, however,
each of
various approaches has its merits and demerits,
or good ``regions'' and bad ones to describe
something (in this case, branes).\footnote{
I would like to thank Taro Tani
for pointing out that.} 
So, thinking of
the results presented above,
we can conclude that
our method is useful
in some aspects to investigate M-theory and String theory.

\parbigskipn

{\Large\bf Acknowledgement}

\parbigskipn
I would like to thank Prof. J. Arafune for careful reading of the
manuscript and useful comments. I am grateful to Akira Matsuda for
many useful discussions and encouragement. 
I would also like to thank Taro Tani for stimulating discussions
and encouragement.
I am obliged to Tsunehide Kuroki for useful comments.

\parbigskipn

\end {document}